\documentclass[12pt]{article}
\usepackage{amsfonts}
\usepackage{amssymb}

\csname @addtoreset\endcsname{equation}{section}
 
\begin{document}
\begin{titlepage}
\begin{flushright}
CERN-TH/2000-141\\ LAPTH-795/2000 \\ hep-th/0005151 
\end{flushright}
\vspace{.5cm} 
\begin{center}
{\Large\bf Superconformal interpretation of BPS states in AdS 
geometries}\\ \vfill {\large  Sergio Ferrara$^\dagger$ and Emery 
Sokatchev$^\ddagger$ }\\ \vfill  \vspace{6pt} $^\dagger$ CERN 
Theoretical Division, CH 1211 Geneva 23, Switzerland 
\\ and 
Department of Physics and Astronomy, University of California, \\ 
Los Angeles, CA 90095, USA 
\\ \vspace{6pt}
$^\ddagger$ Laboratoire d'Annecy-le-Vieux de Physique 
Th\'{e}orique\footnote[1]{UMR 5108 associ{\'e}e {\`a} 
 l'Universit{\'e} de Savoie} LAPTH, Chemin
de Bellevue - BP 110 - F-74941 Annecy-le-Vieux Cedex, France

\end{center}
\vfill 

\begin{center}
{\bf Abstract} 
\end{center}
{\small We carry out a general analysis of the representations of 
the superconformal algebras $\mbox{SU}(2,2/N)$, 
$\mbox{OSp}(8/4,\mathbb{R})$ and $\mbox{OSp}(8^*/2N)$ and give 
their realization in superspace. We present a construction of 
their UIR's by multiplication of the different types of massless 
superfields (``supersingletons"). 
\\ Particular attention is paid to the so-called ``short multiplets". 
Representations undergoing shortening have ``protected dimension" 
and may correspond to BPS states in the dual supergravity theory in 
anti-de Sitter space.\\ These results are relevant for the 
classification of multitrace operators in boundary conformally 
invariant theories as well as for the classification of AdS black 
holes preserving different fractions of supersymmetry.} 
\end{titlepage}

\section{Introduction}

The study of superconformal algebras has recently become of 
central importance because of their dual r\^ole in describing the 
gauge symmetries of supergravity in anti-de Sitter bulk and the 
global symmetries of the boundary field theory \cite{mal,gkp,wit}. 

A special class of configurations which are particularly relevant 
are the so-called BPS states, i.e. dynamical objects corresponding 
to representations which undergo ``shortening". 

These representations can only occur when the conformal dimension 
of a (super)primary operator is ``quantized" in terms of the R 
symmetry quantum numbers and they are at the basis of the 
so-called ``non-renormalization" theorems of supersymmetric 
quantum theories \cite{FIZ}. 

There exist different methods of constructing the UIR's of 
superconformal algebras. One is the so-called oscillator 
construction of the Hilbert space in which a given UIR acts 
\cite{bgg}-\cite{gmz2}. Another one, more appropriate to describe 
field theories, is the realization of such representations on 
superfields defined in superspaces \cite{SS,fwz}. The latter are 
``supermanifolds" which can be regarded as the quotient of the 
conformal supergroup by some of its subgroups. 

In the case of ordinary superspace the subgroup in question is the 
supergroup obtained by exponentiating a non-semisimple 
superalgebra which is the semidirect product of a super-Poincar\'{e} 
graded Lie algebra with dilatation ($\mbox{SO}(1,1)$) and the R 
symmetry algebra. This is the superspace appropriate for non-BPS 
states. Such states correspond to bulk massive states which can 
have ``continuous spectrum" of the AdS mass (or, equivalently, of 
the conformal dimension of the primary fields). 

BPS states are naturally associated to superspaces with lower 
number of ``odd" coordinates and, in most cases, with some 
internal coordinates of a coset space $G/H$. Here $G$ is the R 
symmetry group of the superconformal algebra, i.e. the subalgebra 
of the even part which commutes with the conformal algebra of 
space-time and $H$ is some subgroup of $G$ having the same rank as 
$G$.  

Such superspaces are called ``harmonic" \cite{GIK1} and they are 
characterized by having a subset of the initial odd coordinates 
$\theta$. The complementary number of odd variables determines the 
fraction of supersymmetry preserved by the BPS state. If a BPS 
state preserves $K$ supersymmetries then the $\theta$'s of the 
associated harmonic superspace will transform under some UIR of 
$H_K$. 

For 1/2 BPS states, i.e. states with maximal supersymmetry, the 
superspace involves the minimal number of odd coordinates (half of 
the original one) and $H_K$ is then a maximal subgroup of $G$. On 
the other hand, for states with the minimal fraction of 
supersymmetry $H_K$ reduces to the ``maximal torus" whose Lie 
algebra is the Cartan subalgebra of $G$. 

It is the aim of the present paper to give a comprehensive 
treatment of BPS states related to ``short representations" of 
superconformal algebras for the cases which are most relevant in 
the context of the AdS/CFT correspondence, i.e. the $d=3$ 
($N=8$), $d=6$ and $d=4$ (for arbitrary $N$). The underlying 
conformal field theories correspond to world-volume theories of 
$N_c$ copies of $M_2$, $M_5$ and $D_3$ branes in the large $N_c$ 
limit \cite{AOY}-\cite{ckvp} which are ``dual" to AdS 
supergravities describing the horizon geometry of the branes 
\cite{AGMOO}. 

Some of the results presented in this paper have already appeared 
elsewhere \cite{AFSZ}-\cite{FS3}. \footnote{The new results were 
reported by one of us at the Workshop on "Strings, Branes and 
M-theory" at the CIT-USC Center for Theoretical Physics, Los 
Angeles, California on April 5 and 7, 2000.} Here we give a 
systematic and unified treatment of the BPS states corresponding 
to the three superconformal algebras above. The method we use is 
developed in full detail in the case of the $d=4$ superconformal 
algebra $\mbox{SU}(2,2/N)$ in Sections 2-5. In Section 2 we carry 
out an abstract analysis of the conditions for Grassmann 
(G-)analyticity  \cite{GIO} (the generalization of the familiar 
concept of chirality \cite{fwz}) in a superconformal context. We 
find the  constraints on the conformal dimension and R symmetry 
quantum numbers of a superfield following from the requirement 
that it do not depend on one or more Grassmann variables. 
Introducing G-analyticity in a traditional superspace cannot be 
done without breaking the R symmetry. The latter can be restored 
by extending the superspace by harmonic variables 
\cite{Rosly},\cite{GIK1},\cite{GIK11}-\cite{Bandos} parametrizing 
the coset $G/H_K$. In Section 3 the $(N,p,q)$ harmonic 
superspaces \cite{GIK3,hh} relevant to the description of BPS 
states preserving $p+q/2N$ supersymmetries are reviewed. In 
Section 4 the massless UIR's (``supersingleton" multiplets) 
\cite{ff2}-\cite{bin} of $\mbox{SU}(2,2/N)$ are considered, first 
as constrained superfields in ordinary superspace 
\cite{Siegel,HST} and then, for a part of them, as $(N,p,N-p)$ 
G-analytic harmonic superfields \cite{GIK1,hh}. In Section 5 we 
use supersingleton multiplication to construct UIR's of  
$\mbox{SU}(2,2/N)$. We show that in this way one can reproduce 
the complete classification of UIR's of ref. \cite{dp}. We give 
the full list of BPS states obtained by multiplying chiral and 
G-analytic supersingletons as well as the restricted classes of 
BPS states obtained from one type of G-analytic supersingleton 
alone. We also discuss different kinds of shortening which 
certain superfields (not of the BPS type) may undergo. In 
Sections 6 and 7 we apply the same method to extend these results 
to $d=6$ and $d=3$ for the superalgebras of the maximal 
supersymmetries, i.e., $\mbox{OSp}(8^*/2N)$ and 
$\mbox{OSp}(8/4,\mathbb{R})$. We conclude the paper by listing 
the various BPS states in the physically relevant cases of D3, 
$M_2$ and $M_5$ branes horizon geometry where only one type of 
supersingletons appears. 

Applications of the present results are found \cite{FZa,AFSZ} in 
the classification of multitrace operators in four-dimensional 
$N=4$ $\mbox{SU}(N_c)$ Yang-Mills theory \cite{cs}-\cite{sk}, dual 
to type IIB supergravity on $AdS_5\times S^5$ \cite{mal}.

Another area of interest is the classification of AdS black holes 
\cite{hawk}-\cite{duffl}, according to the fraction of 
supersymmetry preserved by the black hole background.

In a parallel analysis with black holes in asymptotically flat 
background \cite{FMG}, the AdS/CFT correspondence predicts that 
such BPS states should be dual to superconformal states undergoing 
``shortening" of the type discussed here.  

\setcounter{equation}0 
\section{Grassmann analyticity and conformal supersymmetry}\label{GACS}

In this section we shall study the realizations of $D=4$ 
$N$-extended conformal supersymmetry $\mbox{SU}(2,2/N)$ on 
superfields depending on a subset of the $4N$ odd variables. Such 
superfields will be called Grassmann (G-)analytic.

The non-vanishing (anti)commutation relations involving the odd 
generators of the superalgebra $\mbox{SU}(2,2/N)$ are given below: 
\begin{eqnarray}
\big\{Q^i_\alpha, \bar Q_{\dot\alpha j}\big\} &=& 
2\delta^i_j(\sigma^\mu)_{\alpha\dot\alpha}P_\mu\;, \nonumber\\  
\big\{S_{\alpha j} , \bar S^{i}_{\dot\alpha}\big\} &=& 2 
\delta^i_j\;(\sigma^\mu)_{\alpha\dot\alpha}K_\mu~, \nonumber\\ 
\big\{Q^i_\alpha,S^{\beta }_j\big\} &=& - 
\delta^i_j(\sigma^{\mu\nu})_{\alpha }^{\;\;\beta}M_{\mu\nu} - 4 
\delta_{\alpha }^\beta T^i_j - 2\delta_{\alpha }^\beta \delta^i_j 
(R+iD)\;, \nonumber\\ \big[Q^i_\alpha,K_\mu\big] &=& 
-(\sigma_\mu)_{\alpha\dot\alpha}\bar S^{\dot\alpha i}, \ \ \ \ 
\big[\bar Q_{\dot\alpha i}, K_\mu\big] = 
(\sigma_\mu)_{\alpha\dot\alpha}S^\alpha_i, \nonumber\\ 
\big[S_{\alpha i},P_\mu\big] &=& 
-(\sigma_\mu)_{\alpha\dot\alpha}\bar Q^{\dot\alpha}_i, \ \ \ \ 
\big[\bar S^i_{\dot\alpha}, P_\mu\big] = 
(\sigma_\mu)_{\alpha\dot\alpha}Q^{\alpha i}\;, 
\label{2.21}\end{eqnarray} Here the odd generators are 
\footnote{Two-component spinor indices are raised and lowered with 
the help of the Levi-Civita tensor: $\psi^\alpha = 
\epsilon^{\alpha\beta}\psi_\beta$, $\bar\chi^{\dot\alpha} = 
\epsilon^{\dot\alpha\dot\beta}\bar\chi_{\dot\beta}$, $\psi_\alpha 
= \epsilon_{\alpha\beta}\psi^\beta$, $\bar\chi_{\dot\alpha} = 
\epsilon_{\dot\alpha\dot\beta}\bar\chi^{\dot\beta}$; 
$\epsilon_{12} = \epsilon_{\dot 1\dot 2} = -\epsilon^{12} = 
-\epsilon^{\dot 1\dot 2} = 1$
.}: $Q^i_\alpha$,  $\bar Q_{\dot\alpha i} = (Q^i_\alpha)^\dagger$ 
of Poincar\'{e} supersymmetry and $S_{\alpha i}$, $\bar 
S_{\dot\alpha}^i = (S_{\alpha i})^\dagger$ of special conformal 
supersymmetry. The even generators are: $P_\mu$ of translations, 
$K_\mu$ of conformal boosts, $M_{\mu\nu}= -M_{\nu\mu}$ of the 
Lorentz group, $D$ of dilatations, $T^i_j$ of $\mbox{SU}(N)$ and 
$R$ of $U(1)$ (``R charge"). 

Further, the Lorentz and $\mbox{SU}(N)$ generators commute with 
$Q$ as follows: 
\begin{equation}\label{1}
\big[M_{\mu\nu},Q_\alpha\big] = -{1\over 
2}(\sigma_{\mu\nu})_\alpha{}^\beta Q_\beta, \qquad 
\big[M_{\mu\nu},\bar Q_{\dot\alpha}\big] = {1\over 
2}(\tilde\sigma_{\mu\nu})^{\dot\beta}{}_{\dot\alpha} \bar 
Q_{\dot\beta}\;, 
\end{equation}
\begin{equation}\label{2}
  \big[T^i_j, Q^k\big] = \delta^k_j Q^i - {1\over N}\delta^i_j Q^k \; , 
\qquad \big[T^i_j, \bar Q_k\big] = -\delta^i_k \bar Q_j + {1\over 
N}\delta^i_j \bar Q_k \;, 
\end{equation}
and similarly for $S$. Next, the commutators of $Q$ and $S$ with 
the dilatation and R charge generators are given below:
\begin{eqnarray}
  &&[D,Q] = {i\over 2}Q\; ,\ \ \ [D,\bar Q]={i\over 2}\bar Q\;; \nonumber\\
  && [D,S]=-{i\over 2}S\; , \ \ \ [D,\bar S]=-{i\over 2}\bar S\;; 
\label{3}\\
  &&[R,Q] = {4-N\over 2N}Q\; ,\ \ [R,\bar Q]= -{4-N\over 2N}\bar
Q\;;  \nonumber\\
  && [R,S]=-{4-N\over 2N}S\; , \ \ [R,\bar S]={4-N\over 2N}\bar 
S\;. \label{4} 
\end{eqnarray}
Finally, the $\mbox{SU}(N)$ generators $T^i_j$, $(T^i_j)^\dagger = 
T^j_i$, $\sum_{i=1}^NT^i_i=0$ form the algebra
\begin{equation}\label{5}
 [T^i_j, T^k_l] = \delta^k_j T^i_l - \delta^i_l T^k_j\; .
\end{equation}
The rest of the superalgebra $\mbox{SU}(2,2/N)$ is the  conformal 
algebra of $M,P,K,D$ which will not be needed here. 

The superspace traditionally used for the realization of 
$\mbox{SU}(2,2/N)$ (as well as for Poincar\'{e} supersymmetry) is 
given by the real coset 
\begin{equation}
{\mathbb R}^{4\vert 2N,2N} = {\mbox{SU}(2,2/N)\over \{K,S,\bar 
S,M,D,T,R\}} = (x^\mu,\theta^{\alpha}_i, \bar\theta^{\dot\alpha 
i})\;. \label{7.2.2} 
\end{equation}
It is parametrized by 4 even coordinates $x^\mu$ and $2N$ 
left-handed odd spinor coordinates $\theta^{\alpha}_i$ in the 
fundamental of $\mbox{SU}(N)$ together with the $2N$ right-handed 
complex conjugates $\bar\theta^{\dot\alpha i} = 
\overline{\theta^{\alpha}_i}$. The superalgebra is realized on 
superfields $\Phi(x,\theta,\bar\theta)$ defined as functions in 
the coset (\ref{7.2.2}). The generators of the coset denominator 
$K,S,\bar S,M,D,T,R$ act on the superspace coordinates as well as 
on the external indices of the superfield. The latter action is 
given by the matrix parts of these generators, $K_\mu\rightarrow 
k_\mu$, $S_{\alpha i}\rightarrow s_{\alpha i}$, $\bar 
S_{\dot\alpha}^i\rightarrow \bar s_{\dot\alpha}^i$, 
$M_{\mu\nu}\rightarrow  m_{\mu\nu}$, $D\rightarrow i\ell$, 
$T^i_j\rightarrow t^i_j$, $R\rightarrow r$. \footnote{We assign 
the R charge $r_\theta=-{(4-N)/2N}$ to the left-handed Grassmann 
coordinates $\theta^\alpha$ in order to be consistent with the 
convention that chiral superfields $\Phi(\theta)$ have $r=-\ell$ 
for any $N$ (see (\ref{11})). Note that for $N=4$, $r_\theta =0$ 
and the $r$ quantum number becomes a ``central charge" 
\cite{dp,bin}. In this case one refers to the $\mbox{PSU}(2,2/4)$ 
algebra for $r=0$ and to the $\mbox{PU}(2,2/4)$ algebra for 
$r\neq 0$.} According to the definition of a (super)conformal 
primary field, the matrix parts of the transitive generators 
$K,S$ vanish: 
\begin{equation}\label{6}
  s_{\alpha i}\Phi= \bar s_{\dot\alpha}^i\Phi=  k_\mu\Phi= 0
\end{equation}
(the third constraint follows from the first two, see 
(\ref{2.21})). The homogeneous action of the remaining ones, 
$d,l,r,t$, on the superfield and, in particular, on its lowest 
component $\phi(x)=\Phi\vert_{\theta=\bar\theta=0}$ defines the 
latter as an irrep of $\mbox{SO}(1,1)\times \mbox{SL}(2,{\mathbb 
C})\times \mbox{U}(1)\times \mbox{SU}(N)$ with the following 
quantum numbers: 
\begin{equation}\label{7}
  {\cal D}(\ell;j_1,j_2;r;a_1,\ldots, a_{N-1})
\end{equation}
where $\ell$ is the conformal dimension, $j_1,j_2$ are the two 
Lorentz quantum numbers (``spins"), $r$ is the R charge and 
$a_1,\ldots, a_{N-1}$ are the $\mbox{SU}(N)$ Dynkin labels.  
  
\subsection{Chiral superfields}\label{chsup}

The superalgebra $\mbox{SU}(2,2/N)$ can be realized in a smaller 
superspace, called ``chiral" superspace. It is obtained by adding 
half of the Poincar\'{e} supersymmetry generators, for instance, the 
right-handed ones $\bar Q^{\dot\alpha}_i$, to the coset 
denominator: 
\begin{equation}
{\mathbb C}^{4\vert 2N,0} = {\mbox{SU}(2,2/N)\over \{K,S,\bar 
S,M,D,T,R,\bar Q\}} = (x^\mu,\theta^{\alpha}_i)\;. \label{8} 
\end{equation} 
This means adding a new constraint to the set (\ref{6}):
\begin{equation}\label{9}
  \bar q^{\dot\alpha}_i\Phi=0
\end{equation}
where $\bar q$ is the matrix part of the generator $\bar Q$. 
However, in this case the superalgebra (\ref{2.21}) implies 
restrictions on the allowed values of the quantum numbers 
(\ref{7}) \cite{wz}. Indeed, the constraints (\ref{9}), (\ref{6}) 
yield the compatibility condition 
\begin{equation}\label{10}
\{\bar q^{\dot\alpha}_i, \bar s^j_{\dot\beta} \}\Phi = \left[ 
-\delta^j_i (\sigma^{\mu\nu})^{\dot\alpha}{}_{\dot\beta}m_{\mu\nu} 
-2\delta^{\dot\alpha}_{\dot\beta}\left(2t^i_j +\delta^j_i(\ell+r) 
\right)\right] \Phi = 0\;. 
\end{equation}
This is only possible if the superfield (i.e., its first component 
(\ref{7})) carries no right-handed spin, no $\mbox{SU}(N)$ indices 
and has R charge $r=-\ell$: 
\begin{equation}\label{11}
  {\mathbb C}^{4\vert 2N,0} \ \Rightarrow \ 
{\cal D}(\ell;j_1,0;-\ell;0,\ldots,0)\;. 
\end{equation}
Such superfields are called (left-handed) chiral. Note that both 
the superspace (\ref{8}) and the superfields defined in it are 
complex. 
 
Given a general superfield $\Phi(x,\theta,\bar\theta)$, one can 
restrict it to the coset (\ref{8}) by imposing the following 
differential ``chirality" constraint \cite{fwz} 
\begin{equation}\label{12}
  \bar D^{\dot\alpha}_i \Phi(x,\theta,\bar\theta) = 0\;.
\end{equation}
Here $\bar D$ is the right-handed half of the ``covariant spinor 
derivatives" 
\begin{equation}\label{13}
D^i_\alpha = {\partial\over 
\partial \theta^\alpha_i}+
i\bar\theta^{\dot\alpha i}(\sigma^\mu)_{\alpha\dot\alpha} 
\partial_\mu\;, \qquad
\bar D_{\dot\alpha i} = -{\partial\over \partial 
\bar\theta^{\dot\alpha i}}- i\theta^\alpha_i 
(\sigma^\mu)_{\alpha\dot\alpha} \partial_\mu\;. 
\end{equation}
Note that these derivatives are only covariant with respect to the 
super-Poincar\'{e} subalgebra of $\mbox{SU}(2,2/N)$. They obey the 
following anticommutation relations: 
\begin{equation}\label{14}
  \{D^i_\alpha,D^j_\beta\} = \{\bar D_{\dot\alpha i},\bar D_{\dot\beta j}\} =
  0\;, \quad \{D^i_\alpha,\bar D_{\dot\beta j}\} = -2i\delta^i_j
  (\sigma^\mu)_{\alpha\dot\beta}\partial_\mu\;.
\end{equation}
A crucial observation is that the chirality constraint (\ref{12}) 
can be solved by going to the ``left-handed chiral" basis 
\begin{equation}
x^\mu_L=x^\mu+i\theta_{Li}\sigma^\mu\bar\theta^i_L, \ \ \ 
\theta^\alpha_{Li} =\theta^\alpha_i,\ \ \ \bar\theta^{\dot\alpha 
i}_L = \bar\theta^{\dot\alpha i}\;. \label{15}
\end{equation} 
There $\bar D$ becomes just a partial derivative, $\bar 
D_{\dot\alpha i} = -\partial/\partial\bar\theta_L^{\dot\alpha i}$, 
so (\ref{12}) simply implies
\begin{equation}\label{16}
  \Phi=\Phi(x^\mu_L,\theta^\alpha_{Li})\;.
\end{equation}
An important property of the chiral superfields (\ref{16}) is that 
the product of two of them is still a chiral superfield, i.e. they 
form a ``ring structure". Note the close analogy with the typical 
property of ordinary analytic functions. As we shall see in the 
next subsection, this analogy can be further developed. 

\subsection{Grassmann analytic superfields}\label{gan}

A natural question is whether one can find other realizations of 
$\mbox{SU}(2,2/N)$ in superspaces involving only part of the odd 
coordinates. In the chiral case above we chose to add all of the 
right-handed generators $\bar Q^{\dot\alpha}_i$, which form an 
irrep of $\mbox{SU}(N)$, to the coset denominator. Now, let us 
assume for a moment the possibility to break $\mbox{SU}(N)$. 
\footnote{Superspaces of this type can be introduced without 
breaking $\mbox{SU}(N)$ in the framework of harmonic superspace, 
see Section \ref{HSS}.} We can then take just one of the $Q$'s or 
the $\bar Q$'s , e.g., $Q^1_\alpha$ and put it in the denominator. 
The resulting coset has $2N-2$ left-handed and $2N$ right-handed 
odd coordinates: 
\begin{equation}
{\mathbb A}^{4\vert 2N-2,2N} = {\mbox{SU}(2,2/N)\over \{K,S,\bar 
S,M,D,T,R,Q^1\}} = 
(x^\mu,\theta^{\alpha}_2,\ldots,\theta^{\alpha}_N,\bar\theta^1_{\dot\alpha}, 
\ldots, \bar\theta^N_{\dot\alpha}) \;. \label{17} 
\end{equation} 
This means replacing the chirality condition (\ref{9}) by 
\begin{equation}\label{18}
  q^1_\alpha\Phi=0\;.
\end{equation}
Then, a compatibility condition analogous to (\ref{10}) follows 
from the anticommutator 
\begin{equation}\label{19}
 \{q^1_\alpha, s^\beta_1\}\Phi = \left[-(\sigma^{\mu\nu})_\alpha{}^\beta 
m_{\mu\nu} - 2\delta^\beta_\alpha\left(2t^1_1 -\ell +r 
\right)\right]\Phi = 0\;. 
\end{equation}
It implies $(\sigma^{\mu\nu})_\alpha{}^\beta m_{\mu\nu}\Phi = 0$, 
i.e. no left-handed spin, as well as a relation between the 
eigenvalue of the $\mbox{SU}(N)$ generator $t^1_1$, the R charge 
and the conformal dimension: 
\begin{equation}\label{20}
  j_1=0\;, \qquad 2t^1_1 = \ell-r\;.
\end{equation}
Further, anticommuting $q^1_\alpha$ with the remaining projections  
$s^\beta_{2,3,\ldots,N}$, we obtain
\begin{equation}\label{21}
  t^1_i = 0\;, \quad 2\leq i \leq N\;.
\end{equation}

Let us now make a digression and discuss the $\mbox{SU}(N)$ 
generators $t^i_j$. In the Cartan decomposition of the 
$\mbox{SU}(N)$ algebra (\ref{5}) the generators with $1\leq i < j 
\leq N$ are associated to the positive roots (``raising 
operators"). Among them $t^i_{i+1}$, $i=1,\ldots, N-1$ correspond 
to the simple roots, which means that the other raising operators 
are obtained by commuting the simple ones. Similarly, the 
generators with $N\geq i > j \geq 1$ are associated to the 
negative roots (``lowering operators"), the simple ones being 
$t^{i+1}_i$, $i=1,\ldots, N-1$. Finally, the $N-1$ independent 
generators $t^i_i$ (recall that $\sum_{i=1}^Nt^i_i=0$) define the 
$N-1$ charges of the Cartan subalgebra of 
$[\mbox{U}(1)]^{N-1}\subset \mbox{SU}(N)$ as follows: 
\begin{equation}\label{22}
  m_k=t^k_k- t^N_N = t^k_k + {m\over N}\;, 
\qquad 1\leq k \leq N\;, \qquad m= \sum^{N}_{i=1}m_i 
\end{equation}
where $m_N \equiv 0$. An irrep of $\mbox{SU}(N)$ is generated from 
the highest weight state (HWS) $|a_1,\ldots,a_{N-1}\rangle $ 
specified, for example, by the Dynkin labels defined by 
\begin{equation}\label{24}
  a_k=m_k-m_{k+1}\geq 0\;, \qquad 1\leq k \leq N-1\;. 
\end{equation}
Correspondingly, the charges (\ref{22}) of a HWS take eigenvalues 
$m_1\geq m_2 \geq \ldots \geq m_{N-1}\geq m_N=0$. In the language 
of Young tableaux $m_k$ is just the number of boxes in the $k$-th 
row. The HWS is by definition annihilated by all the raising 
operators: 
\begin{equation}\label{23}
  t^i_j|a_1,\ldots,a_{N-1}\rangle =0\;, \quad 1\leq i < j \leq N\;.
\end{equation}
In these terms conditions (\ref{21}) are just a subset of the 
irreducibility conditions (\ref{23}). From (\ref{20}) we obtain 
the following restrictions on the quantum numbers: 
\begin{equation}\label{25}
{2m\over N} - 2m_1 =r-\ell \;.
\end{equation}
 
We can go on and consider a superspace of the type (\ref{17}) 
where the first $p$ $\theta$'s are missing: 
\begin{eqnarray}
{\mathbb A}^{4\vert 2N-2p,2N}  &=& {\mbox{SU}(2,2/N)\over 
\{K,S,\bar S,M,D,T,R,Q^1,\ldots, Q^p\}} \nonumber\\ 
  &=& (x^\mu,\theta^{\alpha}_{p+1},\ldots,\theta^{\alpha}_N,
\bar\theta^1_{\dot\alpha}, 
\ldots, \bar\theta^N_{\dot\alpha}) \;. \label{26} 
\end{eqnarray}
As before, this means to impose
\begin{equation}\label{26'}
  q^i_\alpha\Phi=0\;, \quad 1\leq i \leq p \;. 
\end{equation}
Then, from the anticommutators $\{q^i_\alpha, s^\beta_i\}=0$, 
$1\leq i \leq p$ we obtain conditions similar to (\ref{25}): 
\begin{equation}\label{27}
{2m\over N} - 2m_i =r-\ell\;, \quad 1\leq i \leq p \;. 
\end{equation}
Also, $\{q^i_\alpha, s^\beta_j\}=0$ for $1\leq i < j \leq p$ 
yields a bigger subset of the irreducibility conditions 
(\ref{23}). In addition, this time we obtain a new type of 
condition: 
\begin{equation}\label{28}
  t^i_j|a_1,\ldots,a_{N-1}\rangle =0\;, \quad p\geq i 
> j\geq 1\;.
\end{equation}
The generators in (\ref{28}) are lowering operators of 
$\mbox{SU}(N)$. In fact, these new constraints are corollaries of 
(\ref{27}). Indeed, from (\ref{27}) follows 
\begin{equation}\label{29}
 a_1=\ldots=a_{p-1}=0 \quad \mbox{for}\ p\geq2\;.
\end{equation}
Now, the HWS $|a_1,\ldots,a_{N-1}\rangle$ has the property 
\footnote{The explanation is as follows. The generators 
$t^{k+1}_k$, $t_{k+1}^k$ and $t^k_k-t^{k+1}_{k+1}$ form the 
algebra of $\mbox{SU}(2)_k\subset \mbox{SU}(N)$. The state 
$|a_1,\ldots,a_{N-1}\rangle $ can be regarded as the HWS of an 
irrep of this $\mbox{SU}(2)_k$ of $\mbox{U}(1)$ charge $a_k$, i.e. 
of dimension $a_k+1$. Eq.  (\ref{29'}) then follows from the fact 
that $t^{k+1}_k$ is the lowering operator of $\mbox{SU}(2)_k$.} 
\begin{equation}\label{29'}
  (t^{k+1}_k)^{a_k+1}|a_1,\ldots,a_{N-1}\rangle =0\;.
\end{equation}
Then it is obvious that (\ref{29}) and (\ref{29'}) imply 
(\ref{28}). 

The argument above can be reversed. Take a superfield defined in 
the superspace ${\mathbb A}^{4\vert 2N-2,2N}$ (\ref{17}) whose 
lowest component is in the $\mbox{SU}(N)$ irrep with Dynkin labels 
$[0,\ldots,0,a_p,\ldots,a_{N-1}]$, $p>1$. Then (\ref{28}) holds 
and combining it with the constraint (\ref{18}), we obtain the 
full set of constraints (\ref{26'}). Thus, such a superfield 
effectively lives in a smaller superspace. 

It is clear than we can repeat the same procedure in the 
right-handed sector. This time the starting point will be a 
superspace where $\bar\theta^N_{\dot\alpha}$ is absent (note that 
in our convention $q^1$ and $\bar q_N$ are the HWS's of the 
fundamental irrep of $\mbox{SU}(N)$ and of its conjugate, 
respectively). From the corresponding condition $\bar 
q_N^{\dot\alpha}\Phi=0$ we derive 
\begin{equation}\label{30}
  j_2=0\;, \qquad {2m\over N}=\ell+r\;.
\end{equation}
Going on and removing $q$ right-handed odd variables, 
$\bar\theta^N_{\dot\alpha},\ldots,\bar\theta^{N-q+1}_{\dot\alpha}$, 
i.e., imposing the constraints 
\begin{equation}\label{30'}
  \bar 
q_i^{\dot\alpha}\Phi=0\;, \quad  N-q+1\leq i \leq N \;,
\end{equation}
in addition to (\ref{30}) we find 
\begin{equation}\label{31}
  m_i=0\;, \quad  N-q+1\leq i \leq N-1   \quad \mbox{for}\ q\geq2\;.
\end{equation}
As before, this implies the vanishing of the last $q-1$ Dynkin 
labels:
\begin{equation}\label{32}
  a_i=0\;, \quad  N-q+1\leq i \leq N-1  \quad \mbox{for}\ q\geq2\;.
\end{equation}
Correspondingly, the HWS is annihilated by the lowering operators 
$t^i_j$, $N\geq i > j \geq N-q+1$. 

Finally, we can combine left- and right-handed constraints and 
define the most general G-analytic superspace as follows:
\begin{eqnarray}
 {\mathbb A}^{4\vert 2N-2p,2N-2q} &=& {\mbox{SU}(2,2/N)\over \{K,S,\bar 
S, M,D,T,R,Q^1,\ldots, Q^p,\bar Q_{N-q+1},\ldots, \bar Q_N\}} 
\nonumber\\ 
  &=& (x^\mu,\theta^{\alpha}_{p+1},\ldots,\theta^{\alpha}_N,
\bar\theta^1_{\dot\alpha}, \ldots, 
\bar\theta^{N-q}_{\dot\alpha})\;, \quad p+q\leq N \;.\label{33}  
\end{eqnarray}
Following \cite{hh} we shall call (\ref{33}) an ``$(N,p,q)$ 
superspace" \footnote{The first example of a $(3,2,1)$ superspace 
was given in \cite{GIK3}.}. It is important to realize that 
anticommuting the $Q$'s and $\bar Q$'s in the denominator should 
not produce the translation generator $P_\mu$ which belongs to the 
coset. This explains the condition $p+q\leq N$ in (\ref{33}). The 
superfields defined in this coset are annihilated by a subset of 
the Poincar\'{e} supersymmetry generators: 
\begin{equation}\label{33'}
 q^i_\alpha \Phi = \bar q^{\dot\alpha}_j\Phi=0\;, \qquad 1\leq i 
\leq p\;, \quad N-q+1 \leq j \leq N\;.
\end{equation}
These conditions lead to restrictions on the quantum numbers 
obtained by combining the ones found above: 
\begin{eqnarray}
  &&j_1=j_2=0\;; \nonumber\\
  &&\ell=m_1\;; \nonumber\\
  && r={2m\over N}-m_1\;;\label{34}\\
  && m_1=m_2=\ldots=m_p\;,\nonumber\\
  && m_i=0\;, \quad N-q+1\leq i \leq N-1\;, \quad q\geq 2 \;. \nonumber
\end{eqnarray}
Such $\mbox{SU}(N)$ representations have the first $p-1$ and the 
last $q-1$ Dynkin labels vanishing: 
\begin{equation}\label{35}
  [0,\ldots,0,a_{p},\ldots,a_{N-q},0,\ldots,0]\;.
\end{equation}
   
An interesting limiting case is obtained when $p+q=N$. Such 
superspaces contain exactly one half of the initial number of 
Grassmann variables ($p$ left-handed and $N-p$ right-handed 
spinors). The $\mbox{SU}(N)$ representation of the lowest 
component of the superfield has only one non-vanishing Dynkin 
label, $a_p\neq0$. Consequently, $\ell = a_p$ and 
$r=\left({2p\over N}-1 \right)a_p$. In Section \ref{masup} we 
shall see that in the special case $a_p=1$ such superfields 
describe some of the massless superconformal multiplets. 

We remark that chiral superspace can be viewed as a limiting case 
of the above when, e.g., $p=0$ and $q=N$. In this case only 
$j_1=0$, the other Lorentz quantum number $j_2$ remains arbitrary. 

\setcounter{equation}0 
\section{$(N,p,q)$ harmonic superspace}\label{HSS}

The chiral superspace introduced in Section \ref{chsup} is 
naturally realized in terms of superfields satisfying a 
differential constraint of the type (\ref{12}). The question 
arises if we can formulate similar differential constraints 
restricting a superfield to the G-analytic superspaces of Section 
\ref{gan}. It is quite clear that one should impose constraints 
similar to (\ref{33'}) with the supersymmetry generators replaced 
by spinor covariant derivatives. The only problem is that in 
(\ref{26'}) we have explicitly broken the $\mbox{SU}(N)$ 
invariance, just like when the concept of Grassmann analyticity 
($N=2$) was first introduced in ref. \cite{GIO}. This can be 
repaired by extending the framework of standard superspace to the 
so-called harmonic superspace \cite{GIK1}. 

\subsection{Harmonic variables on the coset $\mbox{SU}(N)/[\mbox{U}(1)]^{N-1}$}

Harmonic superspace is obtained from the ordinary one 
(\ref{7.2.2}) by tensoring it with a coset of the group 
$\mbox{SU}(N)/H$ where $H$ is a maximal subgroup of 
$\mbox{SU}(N)$. In order to be able to describe the most general 
case of G-analytic superfields one has to choose the smallest 
such subgroup, which is the Cartan subgroup 
$[\mbox{U}(1)]^{N-1}$. The resulting coset 
$\mbox{SU}(N)/[\mbox{U}(1)]^{N-1}$ (introduced in \cite{GIK1} for 
$N=2$, \cite{GIK2} for $N=3$ and \cite{Bandos} for arbitrary $N$) 
is a compact complex manifold (``flag manifold" \cite{Knapp,hh}) 
of complex dimension $N(N-1)/2$. Note, however, that $(N,p,q)$ 
superfields for $p\geq 2$ and/or $q\geq 2$ effectively live in 
the smaller cosets $\mbox{SU}(N)/[\mbox{U}(1)]^{N-p-q+1}\times 
\mbox{SU}(p)\times \mbox{SU}(q)$, as we shall explain below (see 
also \cite{hh}). 

\subsubsection{Covariant description of the coset 
$\mbox{SU}(N)/[\mbox{U}(1)]^{N-1}$} 

The harmonic variables $u^I_i$ and their conjugates $u^i_I = 
(u^I_i)^* $  form an $\mbox{SU}(N)$ matrix where $i$ is an index 
in the fundamental representation of $\mbox{SU}(N)$ and 
$I=1,\ldots,N$ are the projections of the second index onto the 
subgroup $[\mbox{U}(1)]^{N-1}$. Further, we define two {\sl 
independent} $\mbox{SU}(N)$ groups, a left one acting on the index 
$i$ and a right one acting on the projected index $I$ of the 
harmonics: 
\begin{equation}\label{3.0}
 (u^I_i)'= \Lambda_i^ju_j^J\Sigma_J^I\;, \qquad \Lambda\in\mbox{SU}(N)_L\;, \quad
\Sigma\in\mbox{SU}(N)_R\;. 
\end{equation}
In particular, the charge operators (\ref{22}) of $\mbox{SU}(N)_R$ 
act on the harmonics as follows: 
\begin{equation}\label{3.1}
m_K\; u^I_i = (\delta_{KI}-\delta_{KN}) u^I_i\;, \qquad m_K\;  
u^i_I = - (\delta_{KI}-\delta_{KN})  u^i_I\;.  
\end{equation}
The harmonics satisfy the following $\mbox{SU}(N)$ defining 
conditions: 
\begin{eqnarray}
 &&u^I_i u^i_J=\delta^I_J~, \nonumber\\ u\in 
\mbox{SU}(N):\quad &&u^I_i u^j_I =\delta^j_i~,\label{3.2}\\ && 
\varepsilon^{i_1\ldots i_N}u^1_{i_1}\ldots u^N_{i_N}=1~. \nonumber 
\end{eqnarray}

\subsubsection{Harmonic functions}
  
A basic assumption of the  harmonic approach to the coset 
$\mbox{SU}(N)/[\mbox{U}(1)]^{N-1}$ is that any harmonic function 
is homogeneous under the action of $[\mbox{U}(1)_R]^{N-1}$, i.e., 
it is an eigenfunction of the charge operators $m_I$, 
\begin{equation}\label{3.3}
m_I\; f^{K_1\ldots K_q}_{L_1\ldots L_r}(u) = (\delta_{K_1I} 
-\delta_{K_1N} - \delta_{L_1I} +\delta_{L_1N} +\ldots)f^{K_1\ldots 
K_q}_{L_1\ldots L_r}(u) 
\end{equation}
(note that the projections (charges) $K_1\ldots K_q;L_1\ldots L_r$ 
are not necessarily all different). Thus the harmonic function 
effectively depends on the $(N^2-1)-(N-1)=N(N-1)$ real coordinates 
of the coset  $\mbox{SU}(N)/[\mbox{U}(1)]^{N-1}$. This description 
of the coset is global and coordinateless. The function 
(\ref{3.3}) is given by its harmonic expansion on the coset (hence 
the term ``harmonic space"). In our $\mbox{SU}(N)$ covariant 
notation this expansion is $[\mbox{U}(1)_R]^{N-1}$ {\sl covariant} 
and $\mbox{SU}(N)_L$ {\sl invariant}. To give a simple example, 
consider the case $N=2$ and the harmonic function 
\begin{eqnarray}
 f^1(u) &=& f^iu^1_{i} + f^{ijk}u^1_i u^1_j u^2_k + \ldots  \nonumber\\
  && + f^{i_1\ldots i_{n+1}j_1\ldots j_{n}}u^1_{i_1}\ldots u^1_{i_{n+1}} 
u^2_{j_1}\ldots u^2_{j_n}+ \ldots \;. \label{3.3''} 
\end{eqnarray}
Note that each term in the expansion has the same overall 
$\mbox{U}(1)_R$ charge $1$. The first coefficient $f^i$ is in the 
fundamental of $\mbox{SU}(2)_L$, and the following ones are 
symmetric in all of their indices (either because $u^1_iu^1_j$ is 
symmetric in $i,j$ or because the antisymmetrization of 
$u^1_iu^2_j$ reduces it to a preceding term in (\ref{3.3''})), 
thus realizing irreps of $\mbox{SU}(2)_L$ of isospin $n+1/2$. As a 
second example, consider the function 
\begin{equation}\label{3.3'''}
  f^1_2(u) \equiv f^{11} = f^{ij}u^1_i u^1_j + f^{ijkl}u^1_i u^1_ju^1_k 
u^2_l + \ldots\;.
\end{equation}
This time the overall charge is even, therefore the irreps of the 
expansion carry integer isospin. 
 
We remark that the irreducible products of harmonics play the r\^{o}le 
of the familiar spherical harmonics in the case $N=2$, where the 
coset $\mbox{SU}(2)/\mbox{U}(1)\sim S^2$ (see \cite{GIK1} for 
details).  

The above $N=2$ examples are generalized to any $N$ as follows. 
\footnote{We are grateful to P. Sorba for help in developing this 
argument.} Consider first a function of the type 
\begin{equation}\label{3.333}
  f^{\stackrel{\underbrace{\mbox{\scriptsize 
1\ldots 1}}}{m_1} \stackrel{\underbrace{\mbox{\scriptsize 2\ldots 
2}}}{m_2}\cdots \stackrel{\underbrace{\mbox{\scriptsize N-1\ldots 
N-1}}}{m_{N-1}}}(u)\;, \qquad { m_1 \geq m_2\geq \ldots \geq 
m_{N-1} } \;. 
\end{equation}
Note that the charges form a sequence corresponding to the 
canonical structure of a Young tableau. This tableau defines the 
smallest irrep of $\mbox{SU}(N)_L$ that one finds in the 
expansion. All the remaining irreps are obtained by the following 
procedure. Denote the HWS of the smallest irrep by its Dynkin 
labels, $|a_1,\ldots,a_{N-1}\rangle$ and that of any irrep present 
in the expansion by $|A_1,\ldots,A_{N-1}\rangle$. The vector 
$|a_1,\ldots,a_{N-1}\rangle$ appears in the multiplet generated by 
the HWS $|A_1,\ldots,A_{N-1}\rangle$, so it can be obtained by the 
action of the lowering operators of $\mbox{SU}(N)_L$: 
\begin{equation}\label{3.334}
  |a_1,\ldots,a_{N-1}\rangle = (t^2_1)^{n_1}(t^3_2)^{n_2}\ldots 
(t^N_{N-1})^{n_{N-1}}|A_1,\ldots,A_{N-1}\rangle \;.
\end{equation}
Here we only use the simple roots; the ordering in (\ref{3.334}) 
is of no importance for our argument. From the $\mbox{SU}(N)$ 
algebra we easily find the following relations between the two 
sets of Dynkin labels:
\begin{equation}\label{3.335}
  A_k = a_k + 2n_k - n_{k-1} - n_{k+1} \geq 0\;, \quad k=1,\ldots, 
N-1\;.
\end{equation}
Note that the coefficients in (\ref{3.335}) form the Cartan matrix 
of $\mbox{SU}(N)$. The total number of boxes of the Young tableaux 
(i.e., number of indices of the coefficients, see below) is given 
by 
\begin{equation}\label{3.335'}
  M = \sum^{N-1}_{k=1}kA_k = m + N n_{N-1}\;.
\end{equation}
Thus one finds an $N-1$-parameter family of irreps where the 
choice of the parameters $n_k$ is only limited by the requirements 
$A_k\geq 0$.  

As an illustration of the above, look at the first term in the 
expansion of the function (\ref{3.333}): 
\begin{equation}\label{3.336}
  f^{i_1\ldots i_{m_1}j_1\ldots j_{m_2}\ldots k_1\ldots 
k_{m_{N-1}}}\; u^1_{i_1}\ldots u^1_{i_{m_1}}  u^2_{j_1}\ldots 
u^2_{j_{m_2}} \ldots u^{N-1}_{k_1}\ldots u^{N-1}_{k_{m_{N-1}}} \;.
\end{equation}
Unlike the simple $\mbox{SU}(2)$ examples above, here the 
coefficients $f$ are not necessarily irreducible under 
$\mbox{SU}(N)_L$. Indeed, they only possess the symmetry 
associated to each type of harmonic projection but no 
antisymmetrization between any two different projections has been 
performed. Comparing the term (\ref{3.336}) to the general case  
(\ref{3.335}) we can say that in (\ref{3.336}) the total number of 
indices (boxes in a Young tableau) is $M=m$, so what is left is 
the $N-2$-parameter family of irreps corresponding to $n_{N-1}=0$. 

The general term in the expansion of the function (\ref{3.333}) is 
obtained from (\ref{3.336}) by multiplying it by the chargeless 
harmonic monomial $u^1_{i_1}\ldots u^N_{i_N}$ (the total 
antisymmetrization of the indices $i_1,\ldots,i_N$ results in an 
$\mbox{SU}(N)_L$ singlet, so it should be eliminated):
$$
f^{\stackrel{\underbrace{\mbox{\scriptsize 1\ldots 1}}}{m_1} 
\stackrel{\underbrace{\mbox{\scriptsize 2\ldots 2}}}{m_2}\cdots 
\stackrel{\underbrace{\mbox{\scriptsize N-1\ldots 
N-1}}}{m_{N-1}}}(u) = 
$$ 
\begin{equation}\label{3.337}
  \sum^\infty_{n_{N-1}=0} f^{i_1\ldots i_M}(u^1)^{m_1+n_{N-1}}\ldots 
(u^{N-1})^{m_{N-1}+n_{N-1}}(u^{N})^{n_{N-1}}\;. 
\end{equation}
We use $n_{N-1}$ from (\ref{3.334}) as the expansion parameter. 
Each term in (\ref{3.337}) has a coefficient with a total number 
of indices $M$ given by (\ref{3.335'}). This coefficient is 
decomposed into a set of $\mbox{SU}(N)_L$ irreps according to the 
rule (\ref{3.335}). 

If the charges ($[\mbox{U}(1)_R]^{N-1}$ projections) of the 
harmonic function do not appear in the canonical order 
(\ref{3.333}), then one should reorder the indices $1,2,\ldots,N$ 
so that they can label a Young tableau. For instance, the $N=4$ 
function $f^{122233}$ should be rewritten as $f^{222331}$, so it 
corresponds to the Young tableau $(3,2,1)$. If a complete set of  
$N$ different projections is present, it can be suppressed, e.g., 
the $N=4$ function $f^{11234}\equiv f^1$. Finally, if the function 
carries lower indices (projections of the complex conjugate 
fundamental representation), they should be converted into sets of 
$N-1$ upper indices, for example,  the $N=4$ function $f^1_4\equiv 
f^{1123}$ or $f^{12}_1\equiv f^{12234}\equiv f^2$.   
 
\subsubsection{Harmonic derivatives}

The harmonic derivatives are operators which  respect the defining 
relations (\ref{3.2}): 
\begin{equation}\label{3.4}
  \partial^{\,I}_J = u^I_i{\partial\over\partial u^J_i} - 
u^i_J{\partial\over\partial u^i_I} - {1\over 
N}\sum^N_{K=1}\delta^{\,I}_J \left(u^K_i{\partial\over\partial 
u^K_i} - u^i_K{\partial\over\partial u^i_K} \right)\;. 
\end{equation} 
They act on the harmonics as follows: 
\begin{equation}
\partial^I_J u^K_i=\delta^K_J u^I_i - {1\over N}\delta^{I}_J u^K_i \;,\qquad
\partial^I_J u^i_K=-\delta^I_K u^i_J +  {1\over N}\delta^{I}_J u^i_K\; .\label{3.5}
\end{equation}
Note that we prefer to treat $u^I_i$ and $u^i_I$ as independent 
variables subject to the constraints (\ref{3.2}).

Clearly, the derivatives $\partial^I_J$ are the generators of the 
group $\mbox{SU}(N)_R$ acting on the $[\mbox{U}(1)_R]^{N-1}$ 
projected indices of the harmonics. The assumption (\ref{3.3}) is 
then translated into the requirement that the harmonic functions 
$f(u)$ are eigenfunctions of the diagonal derivatives 
$\partial^{\,I}_I$ which count the $\mbox{U}(1)_R$ charges: 
\begin{equation}\label{3.6}
(\partial^{\,I}_I-\partial^{\,N}_N) f^{K_1\ldots K_q}_{L_1\ldots 
L_r}(u) = (\delta_{K_1I} -\delta_{K_1N} - \delta_{L_1I} 
+\delta_{L_1N} +\ldots) f^{K_1\ldots K_q}_{L_1\ldots L_r}(u)\;. 
\end{equation}
Then the independent harmonic derivatives on the coset are the 
$N(N-1)/2$ complex derivatives $\partial^{\,I}_J $, $I<J$ 
corresponding to the raising operators of $\mbox{SU}(N)_R$ (or 
their conjugates $\partial^{\,I}_J $, $I>J$ corresponding to the 
lowering operators of $\mbox{SU}(N)_R$). 

{}From the above it follows that the harmonic differential 
conditions 
\begin{equation}\label{a10}
  \partial^{\,I}_J f^{K_1\ldots K_q}_{L_1\ldots L_r}(u) = 0\;, \quad I<J
\end{equation}
impose severe constraints on the harmonic function. Indeed, if the 
function is of the type (\ref{3.333}), it is reduced to just one 
harmonic monomial giving rise to an $\mbox{SU}(N)$ irrep whose HWS 
is labeled by the charges. Any other harmonic function subject to 
the condition (\ref{a10}) must vanish. 

As an example, take $N=2$ and the function $f^1(u)$ (\ref{3.3''}) 
subject to the constraint 
\begin{equation}\label{N2ex}
  \partial^{\,1}_2f^1(u) = 0 \ \Rightarrow \ f^1(u) = f^i u^1_i
\end{equation}
since this is the only term in the expansion (\ref{3.3''}) which 
automatically satisfies the condition (\ref{N2ex}). So, the 
harmonic function is reduced to a doublet of $\mbox{SU}(2)$. 
Similarly, for $N=4$ the function $f^{12}(u)$ is reduced to the 
$\underline 6$ of $\mbox{SU}(4)$. Indeed, the constraints 
$\partial^{\,2}_3 f^{12}(u) = 
\partial^{\,3}_4 f^{12}(u) = 0$ ensure that $f^{12}(u)$ depends on 
$u^1,u^2$ only, $f^{12}(u)= f^{ij}u^1_iu^2_j$. Then the constraint 
$\partial^{\,1}_2 f^{12}(u) =f^{ij}u^1_iu^1_j = 0$ implies $f^{ij} 
= -f^{ji}$. An example of a harmonic function which vanishes if 
subject to the constraint (\ref{a10}) is, e.g., in $N=2$, 
$f_1(u)\equiv f^2(u)$, since no term in its expansion can satisfy 
the condition $\partial^{\,1}_2f^2(u) = 0$.  

Note that not all of the derivatives $\partial^{\,I}_J \;, \ I<J$ 
are independent, as follows from the $\mbox{SU}(N)$ algebra. The 
independent ones,
\begin{equation}\label{rankder}
  \partial^{\,1}_2\;,\ \partial^{\,2}_3\;,\ \ldots,\ \partial^{\,N-1}_N 
\end{equation}
correspond to the simple roots of $\mbox{SU}(N)$. Then the 
constraint (\ref{a10}) is equivalent to 
\begin{equation}\label{a100}
\partial^{\,I}_{I+1} f^{K_1\ldots K_q}_{L_1\ldots L_r}(u) = 0\;, 
\qquad I=1,\ldots, N-1\;. 
\end{equation}

We remark that the coset $\mbox{SU}(N)/\mbox{U}(1)^{N-1}$ can be 
parametrized by $N(N-1)/2$ complex coordinates. In our context 
this amounts to making a choice of the harmonic matrix $u^I_i$ 
such that the group $[\mbox{U}(1)_R]^{N-1}$ is identified with 
$[\mbox{U}(1)_L]^{N-1}\subset \mbox{SU}(N)_L$. Then the harmonic 
derivatives become Cartan's covariant derivatives on the coset. 
The constraints (\ref{a10}) take the form of covariant 
Cauchy-Riemann analyticity conditions. For this reason we can call 
the set of constraints (\ref{a10}) (or (\ref{a100})) harmonic 
(H-)analyticity conditions. The above argument shows that 
H-analyticity is equivalent to defining a HWS of $\mbox{SU}(N)$, 
i.e. it is the $\mbox{SU}(N)$ irreducibility condition on the 
harmonic functions.

\subsection{$(N,p,q)$ harmonic superfields}\label{npqhsf}

The main purpose of introducing harmonic variables is to be able 
to define manifestly $\mbox{SU}(N)$ covariant superfields living 
in the G-analytic superspaces (\ref{33}). This is done following 
the example of the chiral superfields. There we replaced the 
condition (\ref{9}) by the differential chirality constraint 
(\ref{12}). In the case of $(N,p,q)$ analyticity we have to 
replace conditions (\ref{33'}) by analogous differential 
constraints. The crucial point now is to let the superfield depend 
on the harmonic variables and obtain the adequate 
$[\mbox{U}(1)]^{N-1}$ projections with the help of harmonic 
variables: 
\begin{equation}\label{3.7}
 D^I_\alpha \Phi(x,\theta,\bar\theta,u) = 
\bar D^{\dot\alpha}_J\Phi(x,\theta,\bar\theta,u)=0 
\end{equation}
where
\begin{equation}\label{3.8}
D^I_\alpha = D^i_\alpha u_i^I\;, \quad \bar D^{\dot\alpha}_J = 
\bar D^{\dot\alpha}_i u^i_J\;, \quad 1\leq I \leq p\;, \  N-q+1 
\leq J \leq N \;. 
\end{equation}
The derivatives appearing in (\ref{3.7}) anticommute (see 
(\ref{14})), therefore there exists a G-analytic basis in 
superspace, 
\begin{eqnarray}
  &&x^\mu_A =  x^\mu 
-i(\theta_1\sigma^\mu \bar\theta^{1} + \ldots + \theta_p 
\sigma^\mu \bar\theta^{p} - \theta_{N-q+1} \sigma^\mu 
\bar\theta^{N-q+1}-\ldots - \theta_N \sigma^\mu  \bar\theta^{N} 
)\;,\nonumber\\ 
  &&\theta^\alpha_I = \theta^\alpha_i u^i_I\;, \quad 
\bar\theta^{\dot\alpha I} = \bar\theta^{\dot\alpha i} 
u_i^I\;.\label{3.9} 
\end{eqnarray} 
where these derivatives become just $D^I_\alpha = 
\partial/\partial\theta^\alpha_I$, $\bar D_{\dot\alpha\; J} = 
-\partial/\partial\bar\theta^{\dot\alpha\; J}$. Consequently, in 
this basis the analytic superfield (\ref{3.7}) becomes an 
unconstrained function of $N-p$ $\theta$'s and $N-q$ 
$\bar\theta$'s, as well as of the harmonic variables: 
\begin{equation}\label{3.10}
   \Phi(x_A,\theta_{p+1},\ldots,\theta_{N},
\bar\theta^1,\ldots,\bar\theta^{N-q}, u)\;. 
\end{equation} 

Let us now turn to the harmonic dependence in (\ref{3.10}). In 
principle, each component in the $\theta$ expansion of the 
superfield is a harmonic function having an infinite harmonic 
expansion of the type (\ref{3.337}). If we want to deal with a 
finite set of fields, we have to impose a harmonic irreducibility 
condition of the type (\ref{a10}) (or the equivalent subset 
(\ref{a100})). However, in the G-analytic basis (\ref{3.9}) the 
harmonic derivatives become covariant, $D^{\,I}_J$. In particular, 
the derivatives 
\begin{equation}\label{sptn}
  D^{\,I}_{J}= \partial^{\,I}_{J} +2 i\theta_{J}\sigma^\mu 
\bar\theta^{I}\partial_\mu - \theta_J\partial^I + 
\bar\theta^I\bar\partial_J\;, \quad 1\leq I\leq N-q, \ p+1\leq 
J\leq N 
\end{equation}
acquire space-time derivative terms. In the next section we shall 
see that this has important consequences on a G-analytic 
superfield subject to the additional H-analyticity constraints 
\begin{equation}\label{3.11}
  D^{\,I}_{J}\Phi^{[a_1,\ldots,a_{N-1}]}(x_A,\theta_{p+1},\ldots,\theta_{N},
\bar\theta^1,\ldots,\bar\theta^{N-q}, u) = 0\;, \quad 1\leq I < J 
\leq N\;. 
\end{equation}
Here we have indicated the $\mbox{SU}(N)$ representation carried 
by the superfield. 

\subsection{$(N,p,q)$ conformal superfields}\label{npqconf}

So far in this section we have only discussed G-analytic 
superfields as representations of Poincar\'{e} supersymmetry. From the 
analysis of Section \ref{GACS} we know that superconformal 
invariance yields additional restrictions, in particular, on the 
$\mbox{SU}(N)$ irrep carried by the superfield. Adapting the 
arguments of Section \ref{GACS}, one finds that (\ref{3.7}) 
implies the following harmonic conditions (even if we do not 
impose the $\mbox{SU}(N)$ irreducibility conditions (\ref{3.11})): 
$$
D^{\,I}_{I+1}\Phi^{[a_1,\ldots,a_{N-1}]} =  
D^{\,I+1}_{I}\Phi^{[a_1,\ldots,a_{N-1}]} = 0\;, 
$$
\begin{equation}\label{3.12}
1\leq I \leq p-1 \quad \mbox{and} \quad N-q+1\leq I \leq N-1\;. 
\end{equation}
These two subsets of raising and lowering operators of 
$\mbox{SU}(N)$ generate the algebra of $\mbox{SU}(p)\times 
\mbox{SU}(q)$. In the spirit of the coset construction of Section 
\ref{GACS} this means that we have added the factor 
$\mbox{SU}(p)\times \mbox{SU}(q)$ to the denominator of the 
harmonic coset. In other words, a conformally covariant $(N,p,q)$ 
superfield lives not only in a smaller superspace, but also in a 
smaller harmonic space as compared to our initial coset 
$\mbox{SU}(N)/[\mbox{U}(1)]^{N-1}$. From Section \ref{GACS} we 
also know that the Dynkin labels of such a superfield are 
restricted (see (\ref{35})). To summarize, a G-analytic conformal 
superfield has the form 
\begin{equation}\label{3.13'}
\Phi^{[0,\ldots,0,a_p,\ldots,a_{N-q},0,\ldots,0]} 
(x_A,\theta_{p+1},\ldots,\theta_{N}, 
\bar\theta^1,\ldots,\bar\theta^{N-q}, u) 
\end{equation}
and lives in the harmonic coset  
\begin{eqnarray}
  &&{\mbox{SU}(N)\over [\mbox{U}(1)]^{N-p-q+1}\times \mbox{SU}(p)\times 
\mbox{SU}(q)}   \quad \mbox{for}\ p\geq2\;, \ q\geq2\;; 
\nonumber\\ 
  &&{\mbox{SU}(N)\over [\mbox{U}(1)]^{N-q}\times 
\mbox{SU}(q)}   \quad \mbox{for}\ p=0,1\;, \ q\geq2\;; 
\label{3.13}\\ 
  &&{\mbox{SU}(N)\over [\mbox{U}(1)]^{N-p}\times 
\mbox{SU}(p)}   \quad \mbox{for}\ p\geq2\;, \ q=0,1\;; \nonumber\\ 
  &&{\mbox{SU}(N)\over 
[\mbox{U}(1)]^{N-1}}   \quad \mbox{for}\ p=0,1\ \mbox{and} \ 
q=0,1\;. \nonumber
\end{eqnarray}

This effective reduction of the harmonic coset has been pointed 
out in \cite{HL,hh}. For example, in the particular case 
$$
\Phi^{[0,\ldots,0,a_p,0,\ldots,0,a_{N-q},0,\ldots,0]} 
(x_A,\theta_{p+1},\ldots,\theta_{N}, 
\bar\theta^1,\ldots,\bar\theta^{N-q}, u)\quad \Rightarrow 
$$
\begin{equation}\label{3.14}
u\in {\mbox{SU}(N)\over \mbox{S}(\mbox{U}(p)\times 
\mbox{U}(q)\times \mbox{U}(N-p-q))}\;.   
\end{equation}
Note that in the limiting cases $N=p+q$ and $N=p+q+1$ the two 
cosets (\ref{3.13}) and (\ref{3.14}) coincide.

\setcounter{equation}0 
\section{Massless superconformal multiplets}\label{masup}

Massless multiplets are a particular class of superconformal 
multiplets. Their components are fields carrying Lorentz spin 
$(j_1,0)$, $\phi_{\alpha_1\ldots\alpha_{2j_1}}(x)$ or $(0,j_2)$,  
$\bar\phi_{\dot\alpha_1\ldots\dot\alpha_{2j_2}}(x)$ (all indices 
are symmetrized). In addition, they satisfy the massless field 
equations 
\begin{equation}\label{4.0}
\partial^\mu 
\sigma_\mu^{\alpha\dot\alpha}\phi_{\alpha\alpha_2\ldots\alpha_{2j_1}} 
= 0\;, \qquad \partial^\mu 
\sigma_\mu^{\alpha\dot\alpha}
\bar\phi_{\dot\alpha\dot\alpha_2\ldots\dot\alpha_{2j_2}} 
=0 
\end{equation}
(or $\square \phi = 0$ in the case of spin $(0,0)$). These 
massless fields are known \cite{BFH} to form UIR's of the 
conformal algebra $\mbox{SU}(2,2)$ if $\ell=j+1$. Consequently, 
the massless superconformal multiplets form UIR's of 
$\mbox{SU}(2,2/N)$ \cite{dp,bin}. 

In the language of AdS supersymmetry such multiplets are called 
``supersingletons" \cite{NS,GW}. 

In this section we shall formulate the massless multiplets of 
$\mbox{SU}(2,2/N)$ first in terms of ordinary superfields and 
then, for a subclass of them, in $(N,k,N-k)$ harmonic superspace. 
\footnote{The simplest example is provided by the $N=2$ 
hypermultiplet \cite{GIK1}; the next example is the $N=3,4$ 
on-shell SYM field-strength \cite{GIK2}-\cite{Bandos}; the 
generalization to the case $(N,k,N-k)$ was given in \cite{hh}.}

\subsection{Massless multiplets as constrained superfields}

There exist three types of massless $N$-extended superconformal 
multiplets. They can be described in terms of ordinary constrained 
superfields \cite{Siegel,HST}.  

{\it (i).} The first type is given by scalar superfields 
\begin{equation}\label{4.1}
  W^{i_1\ldots 
i_k}(x^\mu, \theta^\alpha_i,\bar\theta^{\dot\alpha i})\;, \qquad 
k=1,\ldots, N-1 
\end{equation}
with $k$ totally antisymmetrized indices of the fundamental 
representation of $\mbox{SU}(N)$ (i.e., carrying Dynkin labels 
$[0,\ldots,0,\stackrel{k}{1},0,\ldots,0]$). They satisfy the 
following constraints: 
\begin{eqnarray}
  &&D^{(j}_\alpha W^{i_1)i_2\ldots i_k}=0\;, \label{1-1}\\
  &&\bar D_{\dot\alpha \{j}W^{i_1\}i_2\ldots i_k}=0  \label{1-2} 
\end{eqnarray}
where $()$ means symmetrization and $\{\}$ means the traceless 
part. In the cases $N=2,3,4$  these constraints define the 
on-shell $N=2$ matter (hyper)multiplet \cite{Sohnius} and the 
$N=3,4$ on-shell super-Yang-Mills multiplets \cite{So}. Their 
generalization to arbitrary $N$ has been given in Refs. 
\cite{Siegel,HST} where it has also been shown that they describe 
on-shell massless multiplets. 

After rewriting the constraints (\ref{1-1}), (\ref{1-2}) in 
harmonic superspace in Section \ref{imass}, we shall see that the 
above massless multiplets are superconformal if 
\begin{equation}\label{4.1.1}
  \ell=1\;, \qquad r ={2k\over N}-1\;.
\end{equation}
We also note their $\mbox{SU}(N)$ quantum numbers
\begin{equation}\label{4.1.2}
  m_1=\ldots=m_k=1\;, \quad m_{k+1}=\ldots=m_{N-1}=0\;, \quad 
m=k\;.
\end{equation}

{\it (ii).} The second type is given by a chiral scalar superfield 
\begin{equation}\label{4.2}
  \bar D_i^{\dot\alpha}\Phi=0
\end{equation}
satisfying the additional constraint (field equation)
\begin{equation}\label{4.3}
  D^{i\;\alpha}D^j_\alpha \Phi=0\;.
\end{equation}
This superfield is an $\mbox{SU}(N)$ singlet. The corresponding 
massless multiplet is superconformal if (see Section \ref{chsup}) 
\begin{equation}\label{4.4}
  \ell=-r=1\;.
\end{equation}

Similarly, one can introduce an antichiral multiplet:
\begin{equation}\label{4.5}
  D^i_\alpha \bar\Phi=0\;, \qquad \bar 
D_{i\;\dot\alpha} D^{\dot\alpha}_j  \bar\Phi=0 
\end{equation}
with quantum numbers
\begin{equation}\label{4.6}
  \ell=r=1\;.
\end{equation}

{\it (iii).} The third type is given by chiral superfields 
carrying external Lorentz spin $(j_1,0)$: 
\begin{equation}\label{4.7}
  \bar D_i^{\dot\alpha} w_{\alpha_1\ldots\alpha_{2j_1}}=0\;.
\end{equation}
Here the $2j_1$ spinor indices are totally symmetrized. These 
superfields are $\mbox{SU}(N)$ singlets. They satisfy the massless 
field equation 
\begin{equation}\label{4.8}
  D^{i\; \alpha}w_{\alpha\alpha_2\ldots\alpha_{2j_1}}=0\;.
\end{equation}
As we have seen in Section \ref{chsup}, conformal supersymmetry 
requires that 
\begin{equation}\label{4.9}
  \ell = -r = j_1+1\;.
\end{equation}

Similarly, one can introduce antichiral superfields with Lorentz 
spin $(0,j_2)$: 
\begin{equation}\label{4.10}
  D^i_\alpha \bar w_{\dot\alpha_1\ldots\dot\alpha_{2j_2}}=0\;, 
\qquad \bar D_i^{\dot\alpha}\bar 
w_{\dot\alpha\dot\alpha_2\ldots\dot\alpha_{2j_2}}=0
\end{equation}
with 
\begin{equation}\label{4.11}
  \ell = r = j_2+1\;.
\end{equation} 

It is straightforward to see that such massless representations 
coincide with the massless supermultiplets of $N$-extended 
Poincar\'e supersymmetry (for an $N=8$ example see ref. 
\cite{gm2}.). 

\subsection{Type (i) massless multiplets as analytic 
superfields}\label{imass} 

Now, let us use the harmonic variables to covariantly project all 
the $\mbox{SU}(N)$ indices in the constraints (\ref{1-1}), 
(\ref{1-2}) onto $[\mbox{U}(1)_R]^{N-1}$. For example, the 
projection 
\begin{equation}\label{4.12}
  W^{12\ldots k} = W^{i_1i_2\ldots i_k}(x,\theta,\bar\theta) u^1_{i_1}u^2_{i_2}\ldots u^k_{i_k}
\end{equation}
satisfies the constraints 
\begin{eqnarray}
  &&D^{1}_\alpha  W^{12\ldots k} = D^{2}_\alpha  W^{12\ldots k} =\ldots =
 D^{k}_\alpha  W^{12\ldots k} = 0\;, \label{4.13}\\
  &&\bar D_{\dot\alpha\; k+1} W^{12\ldots k} = \bar D_{\dot\alpha\; k+2}
W^{12\ldots k} =\ldots = \bar D_{\dot\alpha\; N} W^{12\ldots k} = 
0 \label{4.14} 
\end{eqnarray}
where $D^I_\alpha = D^i_\alpha u_i^I$ and $\bar D_{\dot\alpha\; 
I}= \bar D_{\dot\alpha\; i}u^i_I$. The first of them, eq. 
(\ref{4.13}), is a corollary of the commuting nature of the 
harmonics variables, and the second one, eq. (\ref{4.14}), of the 
defining conditions (\ref{3.2}). In eqs. (\ref{4.13}), 
(\ref{4.14}) one recognizes the conditions for G-analyticity 
(\ref{3.7}) of the type $(N,k,N-k)$. As explained in Section 
\ref{npqhsf}, in the appropriate G-analytic basis (\ref{3.9}) 
$W^{12\ldots k}$ becomes an unconstrained function of $k$ 
$\bar\theta$'s and $N-k$ $\theta$'s: 
\begin{equation}\label{4.16}
   W^{12\ldots k} =  W^{12\ldots k}(x_A,\theta_{k+1},\ldots,\theta_{N},
\bar\theta^1,\ldots,\bar\theta^k, u)\;. 
\end{equation}

It is important to realize that the G-analytic superfield 
(\ref{4.16}) is an $\mbox{SU}(N)$ covariant object only because it 
depends on the harmonic variables. In order to recover the 
original harmonic-independent but constrained superfield $ 
W^{i_1i_2\ldots i_k}(x,\theta,\bar\theta)$ (\ref{1-1}), 
(\ref{1-2}) we need to impose differential constraints involving 
the harmonic variables. In Section \ref{npqhsf} we have shown that 
they take the form of $\mbox{SU}(N)$ irreducibility conditions, 
eq. (\ref{3.11}). In this particular case they are 
\begin{equation}
  \label{a10'}
  D^{\,I}_J W^{12\ldots k} = 0\;, \quad 1\leq I<J \leq N
\end{equation}
or the equivalent set 
\begin{equation}\label{a111}
  D^{\,I}_{I+1} W^{12\ldots k} = 0\;, \quad 1\leq I<J \leq N-1\;. 
\end{equation}
In the initial real basis (\ref{7.2.2}) of the full superspace 
${\mathbb R}^{4\vert 2N,2N}$ these constraints simply mean that 
the superfield is a polynomial in the harmonics, as in 
(\ref{4.12}). However, in the G-analytic basis (\ref{3.9}) the 
harmonic derivatives (\ref{sptn}) contain space-time derivatives. 
This leads to a number of constraints on the component fields. The 
detailed analysis can be found in \cite{FS1}, here we only recall 
the final result: 
\begin{eqnarray}
 W^{12\ldots k} &=&\phi^{12\ldots k} \nonumber\\
 &&+\bar\theta^1_{\dot\alpha}\bar\psi^{\dot\alpha\; 23\ldots k} + 
\ldots + \bar\theta^k_{\dot\alpha}\bar\psi^{\dot\alpha\; 12\ldots 
k-1} \nonumber\\ 
 &&+ \theta^\alpha_{k+1}\chi_{\alpha}^{1\ldots k\; k+1}  + \ldots + 
\theta^\alpha_{N}\chi_{\alpha}^{1\ldots k\; N} \nonumber\\ 
 &&+\bar\theta^1_{\dot\alpha} \bar\theta^2_{\dot\beta} 
\bar\psi^{(\dot\alpha\dot\beta)\; 3\ldots k} + \ldots + 
\bar\theta^{k-1}_{\dot\alpha} \bar\theta^k_{\dot\beta} 
\bar\psi^{(\dot\alpha\dot\beta)\; 1\ldots k-2}\nonumber\\ 
 && + \theta^\alpha_{k+1} \theta^\beta_{ k+2} 
\chi_{(\alpha\beta)}^{1\ldots k\; k+1\; k+2} 
 + \ldots + \theta^\alpha_{N-1} \theta^\beta_{N} 
\chi_{(\alpha\beta)}^{1\ldots k\; N-1\; N}\nonumber\\ 
 &&  \ldots\nonumber\\ 
 && + \bar\theta^1_{\dot\alpha_1}\ldots \bar\theta^k_{\dot\alpha_k} 
\bar\psi^{(\dot\alpha_1\ldots\dot\alpha_k)} 
+\theta^{\alpha_1}_{k+1} \ldots \theta^{\alpha_{N-k}}_{N} 
\chi_{(\alpha_1\ldots\alpha_{N-k})} \nonumber\\ 
 && + \mbox{derivative terms}\ . \label{expan} 
\end{eqnarray}
Here all the component fields belong to totally antisymmetric 
irreps of $\mbox{SU}(N)$, e.g., $\phi^{12\ldots k}(x,u) = 
\phi^{[i_1i_2\ldots i_k]}(x) u^1_{i_1} u^2_{i_2}\ldots u^k_{i_k}$. 
Further, these fields satisfy massless field equations of the type 
(\ref{4.0}). 

We conclude this section by a remark concerning the conformal 
properties of the above multiplets. The $(N,k,N-k)$ analytic 
superfield $W^{12\ldots k}$ is characterized by the $\mbox{SU}(N)$ 
quantum numbers $m_1=\ldots=m_k=1,\ m_{k+1}=\ldots=m_{N-1}=0$. 
{}From eqs. (\ref{34}) we see that if 
\begin{equation}\label{rch}
 \ell_k=1\;, \qquad r_k={2k\over N}-1
\end{equation}
$W^{12\ldots k}$ realizes a massless UIR of the superconformal 
algebra. 

\setcounter{equation}0 
\section{UIR's of $D=4$ $N$-extended conformal 
supersymmetry}\label{sect5} 

In this section we shall show how the complete classification of 
UIR's of $\mbox{SU}(2,2/N)$ found in \cite{dp} (see also 
\cite{mss}, \cite{bin} for the massless case) can be obtained by 
multiplying the three types of massless superfields introduced in 
Section \ref{masup}. 

\subsection{The three series of UIR's}\label{3ser} 

The results of \cite{dp} \footnote{Our conventions differ from 
those of \cite{dp} in the following sense: $r\rightarrow -r$, 
${2m/ N}\rightarrow 2m_1-{2m/ N}$.} fall into three distinct 
series. The simplest one (called series C in \cite{FS1}) is given 
by the following conditions: 
\begin{equation}\label{C}
  {\rm C)}\qquad \ell = m_1\;, \quad r={2m\over N}-m_1\;, \quad j_1=j_2=0\;. 
\end{equation} 

We can construct the superfield realization of series C by 
multiplying massless G-analytic superfields \footnote{Series of 
operators obtained as powers of the $N=4$ super-Yang-Mills field 
strength considered as a G-analytic harmonic superfield were 
introduced in \cite{HWest}. They were identified with short 
multiplets of $SU(2,2/4)$ and their correspondence with the K-K 
spectrum of IIB supergravity was established in \cite{AF}.} 
(``supersingletons") of the type (\ref{4.16}): 
\begin{equation}\label{5.1}
 W^{[a_1,\ldots,a_{N-1}]} = (W^1)^{a_1}(W^{12})^{a_2}\ldots 
(W^{12\ldots N-1})^{a_{N-1}}\;.
\end{equation}
Since each factor in (\ref{5.1}) satisfies the usual harmonic 
irreducibility constraints, the same is true for the product: 
\begin{equation}\label{5.2}
  D^I_J W^{[a_1,\ldots,a_{N-1}]} = 0\;, \qquad 1\leq J < I \leq 
N\;.
\end{equation}
As a result, the lowest component of the superfield (\ref{5.1}) is 
an irrep of $\mbox{SU}(N)$ with Dynkin labels 
$[a_1,\ldots,a_{N-1}]$. This is easily seen by realizing that: i) 
all the $\mbox{SU}(N)$ indices projected with harmonics $u^K_i$ 
for a given $K$ are symmetrized; ii) their total number is 
$m_K=\sum^{N-1}_{i=K}a_i$; iii) the harmonic conditions 
(\ref{5.2}) remove all symmetrizations between indices projected 
with different harmonics $u^K_i$ and $u^L_i$. All this reproduces 
the structure of a Young tableau with numbers of boxes 
$(m_1,m_2,\ldots,m_{N-1})$, i.e. Dynkin labels 
$[a_1,\ldots,a_{N-1}]$. 

Further, from (\ref{rch}) we find $\ell= \sum^{N-1}_{k=1} 
a_k\ell_k = m_1$ and $r=\sum^{N-1}_{k=1} a_k r_k = {2m\over 
N}-m_1$, which exactly reproduces (\ref{C}). Thus, we have proved 
that the complete series C is realized by the product (\ref{5.1}) 
of massless multiplets. 

We remark that for a generic choice of the Dynkin labels the 
superfield (\ref{5.1}) is $(N,1,1)$ G-analytic. However, if the 
first $p-1$ or the last $q-1$ (or both) factors in (\ref{5.1}) are 
absent, i.e., if the corresponding Dynkin labels vanish, we obtain 
further analyticity conditions of the type $(N,p,q)$, in accord 
with (\ref{3.13'}). We should mention that in ref. \cite{dp} a 
list of the possible superconformal differential conditions on 
superfields is given. There one only finds $(N,1,1)$ G-analyticity 
conditions, but this can be explained by the above observation.   

The second series (called B in \cite{FS1}) is given by the 
following conditions:  
\begin{equation}\label{B}
  {\rm B)}\qquad \ell = -r+ {2m\over N} \geq 2+2j_1+r+2m_1 -{2m\over 
N}\;, \quad j_2=0
\end{equation}
(or  $j_1\rightarrow j_2$, $r\rightarrow -r$, ${2m\over 
N}\rightarrow 2m_1-{2m\over N}$). It can be obtained by 
multiplying the G-analytic massless superfield (\ref{5.1}) by 
left-handed chiral ones as follows: 
\begin{equation}\label{5.3}
  w_{\alpha_1\ldots\alpha_{2j_1}}\;\Phi^k\; W^{[a_1,\ldots,a_{N-1}]}
\end{equation}
where $k\geq 0$ is an integer. The first factor in (\ref{5.3}) 
brings in the Lorentz spin $(j_1,0)$. The second factor adjusts 
the dimension and R charge of the series, 
\begin{equation}\label{5.4}
\ell = 1+j_1+m_1+k\;, \qquad r=-1-j_1-k - m_1 + {2m\over N}\;, 
\end{equation}
so that they exactly match (\ref{B}). The conformal bound in  
(\ref{B}) is obtained for $k=0$, i.e. without employing any scalar 
chiral superfields. The alternative series of this type is 
obtained by replacing chiral by antichiral superfields. 

Finally, the most general series (called A in \cite{FS1}) is 
given by the following conditions: 
\begin{equation}\label{A}
  {\rm A)}\qquad   \ell \geq  2+2j_2-r+{2m\over N}\geq 2+2j_1+r+2m_1-{2m\over 
N}  
\end{equation}
(or  $j_1\rightarrow j_2$, $r\rightarrow -r$, ${2m\over 
N}\rightarrow 2m_1-{2m\over N}$).
This series is obtained by multiplying together all possible types 
of  massless superfields:   
\begin{equation}\label{5.6}
  w_{\alpha_1\ldots\alpha_{2j_1}}\;\bar 
w_{\dot\alpha_1\ldots\dot\alpha_{2j_2}}\; \Phi^k\; \bar\Phi^s\;  
W^{[a_1,\ldots,a_{N-1}]} 
\end{equation}
where $k\geq s\geq 0$ are integers. This time we find
\begin{equation}\label{5.7}
  \ell = 2+j_1+j_2+m_1+k+s\;, \qquad r=j_2-j_1-k+s-m_1+{2m\over 
N} 
\end{equation}
which corresponds to (\ref{A}). The two conformal bounds in  
(\ref{A}) are saturated for $s=0$ or $k=s=0$, i.e. without 
employing one or the other type (or both) of scalar chiral 
superfields. These bounds correspond to superfields satisfying 
differential constraints, as explained in Section 5.3. The 
alternative series is obtained by taking $s\geq k \geq 0$. 

Note that in the abstract series (\ref{B}) and (\ref{A}) the 
dimension $\ell$ and R charge $r$ can be any real numbers. In 
order to account for this, the powers $k$ and $s$ in (\ref{5.3}) 
and (\ref{5.6}) will have to take non-integer values, although 
this might violate unitarity. This does not happen for series C 
where $\ell$ is always integer and $r$ is rational. 

One final remark concerns the unitarity of the above series of 
representations. Earlier we mentioned that the massless multiplets 
(supersingletons) are known to be UIR's of the superconformal 
algebra. Then it is clear that by multiplying them as we did above 
we automatically obtain series of UIR's. 

\subsection{Series obtained from one type of supersingleton}

In Section \ref{3ser} we used all possible G-analytic 
supersingletons $W^{12\ldots n}$ with $1\leq n\leq N-1$ to 
reproduce the complete series C. An alternative approach is to 
use different realizations of the same type of supersingleton 
(i.e., for a fixed value of $n$). We presented a similar 
construction in \cite{FS1}, where we only considered the case 
$n=N/2$ (for even $N$). The generalization is straightforward. The 
result is a series of UIR's which is a particular case of the 
series B above. 

The supersingleton $W^{12\ldots n}$ can be equivalently rewritten 
by choosing different harmonic projections of its $\mbox{SU}(N)$ 
indices and, consequently, different sets of G-analyticity 
constraints. This amounts to superfields of the type  
\begin{equation}\label{5.8}
  W^{I_1I_2\ldots I_n}
(\theta_{J_{n+1}},\ldots, \theta_{J_N}, 
\bar\theta^{I_1},\ldots,\bar\theta^{I_n}) 
\end{equation}
where $I_1,\dots,I_n$ and $J_{n+1},\dots,J_N$ are two 
complementary sets of $N$ indices. Each of these superfields 
depends on $2N$ Grassmann variables, i.e. half of the total number 
of $4N$. This is the minimal size of a G-analytic superspace, so 
we can say that the $W$'s are the ``shortest" superfields 
(superconformal multiplets). 

The idea now is to start multiplying different versions of the 
$W$'s of the type (\ref{5.8}) (for a fixed value of $n$) in order 
to obtain composite objects depending on various numbers of odd 
variables. The following choice of $W$'s and of the order of 
multiplication covers all possible intermediate types of 
G-analyticity: 
\begin{eqnarray}
  &&A(p_1,p_2,\ldots,p_{N-1})\nonumber\\
  && = [W^{1\ldots n}(\theta_{n+1 \ldots  N} 
\bar\theta^{1  \ldots  n})]^{p_1 + \ldots +p_{N-1}}\nonumber\\ 
  &&\times [W^{1\ldots n-1\; 
n+1}(\theta_{\underline{n}\;{n+2} \ldots  N} \bar\theta^{1\ldots 
n-1\; \underline{n+1}}) ]^{p_2 + \ldots +p_{N-1}}\nonumber\\ 
  &&\times [W^{1\ldots n-1\; 
n+2}(\theta_{n\;  {n+1}\; {n+3} \ldots N}  \bar\theta^{1\ldots 
n-1\; \underline{n+2}} ) ]^{p_3 + \ldots +p_{N-1}}\nonumber\\ 
  && \cdots     \nonumber\\ 
  &&\times [W^{1\ldots n-1\; 
N-1}(\theta_{n \ldots {N-2} \; N} \bar\theta^{1\ldots n-1\; 
\underline{N-1}})]^{p_{N-n} + \ldots +p_{N-1}}\nonumber\\ 
  &&\times [W^{1\ldots n-2\; n\; n+1}(\theta_{\underline{n-1}\; {n+2} 
\ldots  N} \bar\theta^{1\ldots n-2\; n\; n+1})]^{p_{N-n+1} + 
\ldots +p_{N-1}}\nonumber\\ 
  &&\times [W^{1\ldots n-3\; n-1\; n\; n+1}(\theta_{\underline{n-2}\; n+2
 \ldots N} 
\bar\theta^{1\ldots n-3\; n-1\; n\; n+1}) ]^{p_{N-n+2} + \ldots 
+p_{N-1}}\nonumber\\ 
  && \cdots     \nonumber\\ 
  &&\times [W^{13\ldots n+1}(\theta_{\underline{2}\; n+2 
 \ldots N} 
\bar\theta^{13\ldots n+1}) ]^{p_{N-2}+p_{N-1}} \nonumber\\  
  &&\times [W^{23\ldots n+1}(\theta_{\underline{1}\; {n+2} \ldots  N} 
\bar\theta^{23\ldots n+1}) ]^{p_{N-1}} \;.         
\label{verylong} 
\end{eqnarray}

The power $\sum^{N-1}_{r=k} p_r$ of the $k$-th $W$ is chosen in 
such a way that each new $p_r$ corresponds to bringing in a new 
realization of the same supersingleton. As a result, at each step 
a new $\theta$ or $\bar\theta$ appears (they are underlined in 
(\ref{verylong})), thus adding new odd dimensions to the 
G-analytic superspace. The only exception of this rule is the 
second step at which both a new $\theta$ and a new $\bar\theta$ 
appear. So, the series (\ref{verylong}) covers the cases 
$(N,n,N-n)$, $(N,n-1,N-n-1)$ and  then all intermediate cases up 
to $(N,1,0)$. 

The superfield $A(p_1,p_2,\ldots,p_{N-1})$ should be submitted to 
the same H-analyticity constraints as one would impose on 
$W^{1\ldots n}$ alone, 
\begin{equation}\label{alone}
  D^{\, I}_{I+1}A(p_1,p_2,\ldots,p_{N-1})=0\;, \qquad 
I=1,2,\ldots, N-1\;. 
\end{equation}
This is clearly compatible with the G-analyticity conditions on 
$A(p_1,p_2,\ldots,p_{N-1})$ since they form a subset of these on 
$W^{1\ldots n}$. As before, H-analyticity makes 
$A(p_1,p_2,\ldots,p_{N-1})$ irreducible under $\mbox{SU}(N)$. 

By counting the number of occurrences of each projection 
$1,2,\ldots,N-1$ and the dimensions and R charges in 
(\ref{verylong}), we easily find the relations 
\begin{equation}\label{m-l}
   \ell= \sum^{N-1}_{k=1}kp_k\;, \quad m_1= \ell-p_{N-1}\;, \quad m=n\ell\;,
\quad  r= \left({2n\over N}-1 \right)\ell\;. 
\end{equation}
If $N=2n$ this series has no R charge. If $p_{N-1}=0$ the product
(\ref{verylong}) represents a G-analytic superfield and is thus a 
particular case of the series C. If $p_{N-1}\geq 1$ it depends on 
all $\theta$'s and on all $\bar\theta$'s but $\bar\theta^N$, so it 
is a particular case of the series  B (\ref{5.4}) with $j_1=0$. 

Finally, the Dynkin labels of the $\mbox{SU}(N)$ irrep carried by 
the first component of $A(p_1,p_2,\ldots,p_{N-1})$ are given 
below: 
\begin{eqnarray}
  && a_{1}=p_{N-2} \;,  \nonumber\\
  && a_{2}= p_{N-3}\;, \quad  \ldots\;, \quad  a_{n-2}= p_{N-n+1}\;,  \nonumber\\
  && a_{n-1}= (N-n-2)\sum_{k=N-n+1}^{N-1} p_k + 
  \sum_{k=2}^{N-n} (k-1)p_{k}\;, \nonumber\\ 
  && a_{n}= p_{1} \;,  \label{DL}  \\
  &&  a_{n+1}= p_2 + \sum_{k=N-n+1}^{N-1} (k-N+n)p_{k}\;,   \nonumber\\
  && a_{n+2}= p_{3} \;, \quad  \ldots\;, \quad a_{N-2} = 
p_{N-n-1}\;,  \nonumber\\ 
  && a_{N-1}= \sum_{k=N-n}^{N-1} p_k\;.   \nonumber
\end{eqnarray}  

An interesting particular case is obtained if $a_{N-1}= 0$. This 
implies $p_{N-n}=\ldots=p_{N-1}=0$, so $a_1=\ldots=a_{n-2}=0$. In 
other words, this is a G-analytic superfield of the type 
$(N,n-1,2)$. The remaining Dynkin labels are 
$a_{n-1}=\sum_{k=2}^{N-n-1} (k-1)p_{k} $, $a_n=p_1$, 
$a_{n+1}=p_2$, \ldots, $a_{N-2}=p_{N-n-1}$. In general, none of 
these labels vanishes, therefore the harmonic coset in which this 
$(N,n-1,2)$ superfield lives is not smaller than the expected one, 
$\mbox{SU}(N)/[\mbox{U}(1)]^{N-n}\times \mbox{SU}(n-1)\times 
\mbox{SU}(2)$.   

\subsection{Shortness conditions}\label{shortening}

In the AdS literature the term ``short" applies to multiplets 
which do not reach their maximal spin (equal to $(j_1 + {N\over 
2}, j_2 + {N\over 2})$ where $(j_1,j_2)$ is the spin of the first 
component) or which contain constrained fields like, e.g., 
conserved vectors. Our construction of the UIR's of 
$\mbox{SU}(2,2/N)$ in terms of supersingletons allows us to easily 
find out when and what type of ``shortness" condition takes place. 

To this end we recall that the building blocks $w$, $\Phi$ and $W$ 
are all constrained superfields corresponding to the ``ultrashort" 
supersingleton multiplets. They are either G-analytic 
((\ref{4.13}), (\ref{4.14})) or chiral ((\ref{4.2}), (\ref{4.7})). 
In addition, they satisfy on-shell constraints which take the form 
of $\mbox{SU}(N)$ irreducibility harmonic conditions (\ref{a10'}) 
in the G-analytic case or are of the type (\ref{4.3}) or 
(\ref{4.8}) in the chiral case. 

Now, the most general product of chiral, antichiral and G-analytic 
superfields as in the series A (\ref{5.6}) only satisfies the 
harmonic constraints (\ref{a10'}) (recall that $w$ and $\Phi$ are 
harmonic-independent). However, there is a number of particular 
cases where some constraints on the $\theta$ dependence still take 
place. 

i) The product 
$w_{\alpha_1\ldots\alpha_{2j_1}}\;W^{[a_1,\ldots,a_{N-1}]}$  
satisfies the intersection of the constraints (\ref{4.7}), 
(\ref{4.8}) of the factor $w$ with the G-analyticity ones of the 
factor $W$. In the generic case the latter is of the type 
$(N,1,1)$, so we have 
\begin{eqnarray}
  &&\bar D^{\dot\alpha}_N 
(w_{\alpha_1\ldots\alpha_{2j_1}}\;W^{[a_1,\ldots,a_{N-1}]}) = 0\;, 
\label{5.9}\\ 
  && D^{1\alpha}(w_{\alpha\alpha_2\ldots\alpha_{2j_1}}\;W^{[a_1,\ldots,a_{N-1}]}) = 0\;.
 \label{5.10}
\end{eqnarray}
If $W$ carries Dynkin labels like in (\ref{3.13'}), it is of the 
type $(N,p,q)$ and, correspondingly, we obtain $q$ equations like 
(\ref{5.9}) and $p$ ones like (\ref{5.10}).

Similarly, the product $\Phi\;W^{[a_1,\ldots,a_{N-1}]}$ 
satisfies the constraints 
\begin{eqnarray}
  &&\bar D^{\dot\alpha}_N 
(\Phi\;W^{[a_1,\ldots,a_{N-1}]}) = 0\;, \label{5.11}\\ 
  && D^{1\alpha}D^1_\alpha(\Phi\;W^{[a_1,\ldots,a_{N-1}]}) = 0
 \label{5.12}
\end{eqnarray}
or more of the same type is $W$ is $(N,p,q)$ analytic.

ii) The bilinear products of chiral with anti-chiral 
superfields are current-like objects.
They satisfy constraints which turn  the top 
spin in the superfield into a conserved ``current". 
The simplest example is the bilinear $\Phi\bar \Phi$: 
\begin{equation}\label{5.13}
  D^{i\alpha}D^j_\alpha(\Phi\bar \Phi) = 0\;,\qquad \bar D_{i\dot\alpha} \bar 
D_j^{\dot\alpha}(\Phi\bar \Phi) = 0\;. 
\end{equation}
These constraints can be weakened if we multiply $\Phi\bar \Phi$ 
by a G-analytic factor $W$. In this case only certain projections 
of (\ref{5.13}) are preserved, e.g., 
\begin{equation}\label{5.14}
  D^{1\alpha}D^1_\alpha(\Phi\bar \Phi\;W^{[a_1,\ldots,a_{N-1}]}) = 
\bar D_{N\dot\alpha} \bar 
D_N ^{\dot\alpha}(\Phi\bar \Phi\;W^{[a_1,\ldots,a_{N-1}]}) = 0\;. 
\end{equation}

Yet another current-like object is the bilinear 
$w_{\alpha_1\ldots\alpha_{2j_1}}\;\bar 
w_{\dot\alpha_1\ldots\dot\alpha_{2j_2}}$. It satisfies the 
constraints
\begin{eqnarray}
  &&\bar D^{\dot\alpha}_i 
(w_{\alpha_1\ldots\alpha_{2j_1}}\;\bar 
w_{\dot\alpha\dot\alpha_2\ldots\dot\alpha_{2j_2}}) = 0\;, 
\label{5.15}\\ 
  && D^{i\alpha}(w_{\alpha\alpha_2\ldots\alpha_{2j_1}}\;\bar 
w_{\dot\alpha_1\ldots\dot\alpha_{2j_2}}) = 0\;. 
 \label{5.16}
\end{eqnarray} 
As before, the product $w_{\alpha_1\ldots\alpha_{2j_1}}\;\bar 
w_{\dot\alpha_1\ldots\dot\alpha_{2j_2}}\;W^{[a_1,\ldots,a_{N-1}]}$ 
satisfies only the corresponding projections of the above.

Similarly, the bilinear $w_{\alpha_1\ldots\alpha_{2j_1}}\;\bar\Phi$ 
satisfies the constraints 
\begin{eqnarray}
  && D^{i\alpha}(w_{\alpha\alpha_2\ldots\alpha_{2j_1}}\;\bar 
\Phi)=0\;, \label{5.17}\\ 
  &&\bar D_{i\dot\alpha} \bar 
D_j^{\dot\alpha}(w_{\alpha_1\ldots\alpha_{2j_1}}\;\bar \Phi)=0\;. 
\label{5.18} 
\end{eqnarray}

iii) A different class of ``short" objects are obtained from the 
most general product (\ref{5.6}) of series A either by setting 
$s=0$ or $j_2=0$ and $s=1$. In other words, we take  the 
current-like bilinears above and multiply them by a BPS object 
(i.e., product of a chiral and a G-analytic factors). The 
resulting objects satisfy the constraints (for a generic $W$): 
\begin{eqnarray}
  && \bar D^{\dot\alpha}_N(w_{\alpha_1\ldots\alpha_{2j_1}}\;\bar 
w_{\dot\alpha\dot\alpha_2\ldots\dot\alpha_{2j_2}}\; \Phi^k\;  
W^{[a_1,\ldots,a_{N-1}]})=0\;, \label{5.17'}\\ 
  &&\bar D_{N\dot\alpha} \bar 
D_N ^{\dot\alpha}(w_{\alpha_1\ldots\alpha_{2j_1}}\;\bar\Phi\; \Phi^k\;  
W^{[a_1,\ldots,a_{N-1}]})=0\;. 
\label{5.18'} 
\end{eqnarray} 
We call such objects ``intermediate short". Note that they 
saturate the first conformal bound in (\ref{A}). Intermediate 
short multiplets, as they are defined above, will also occur in 
$d=6$ and $d=3$ (see Sections \ref{short6} and \ref{short3}).

\subsection{BPS states of $\mbox{SU}(2,2/N)$}

Here we give a summary of the  $\mbox{SU}(2,2/N)$ multiplets which 
correspond to BPS states. \footnote{Note that such BPS states have 
a close resemblance to BPS Poincar\'e multiplets in five 
dimensions \cite{HULL}, as expected by a limiting procedure.} They 
are realized in terms of superfields which do not depend on at 
least one spinor coordinate. There are three distinct ways to 
obtain such multiplets. 

\subsubsection{$(p,q)$ BPS states} 

Superfields which do not depend on the first $p$ $\theta$'s and 
the last $q$ $\bar\theta$'s are obtained by multiplying 
G-analytic objects: 
$$
{p+q\over 2N}\mbox{ BPS:}\qquad 
W^{[0,\ldots,0,a_p,a_{p+1},\ldots,a_{N-q},0,\ldots,0]} 
(\theta_{p+1},\ldots,\theta_{N}, 
\bar\theta^1,\ldots,\bar\theta^{N-q}) 
$$
\begin{equation}\label{5.19}
=(W^{12\ldots p})^{a_p}(W^{12\ldots p+1})^{a_{p+1}}\ldots 
(W^{12\ldots N-q})^{a_{N-q}}   
\end{equation}
where
\begin{equation}\label{5.20}
 1\leq p,q\leq N-1\;, \qquad  p+q\leq N\;.
\end{equation}
Note that the fraction of supersymmetry preserved by a $(p,q)$ BPS 
state ranges as follows: 
\begin{equation}\label{5.21'}
{1\over N} \leq {p+q\over 2N} \leq {1\over 2}\;.
\end{equation}
The two end points are obtained for $p=q=1$ and for $p+q=N$.

Such states have the first $p-1$ and the last $q-1$ $\mbox{SU}(N)$ Dynkin 
labels vanishing. The remaining quantum numbers are: 
\begin{equation}\label{5.21}
  \ell = \sum_{k=p}^{N-q}a_k\;, \quad j_1=j_2=0\;, \quad r = 
\sum_{k=p}^{N-q}({2k\over N}-1)a_k\;. 
\end{equation}
Generically, such superfields live in the harmonic space
\begin{equation}\label{5.22}
{\mbox{SU}(N)\over [\mbox{U}(1)]^{N-p-q+1}\times 
\mbox{SU}(p)\times \mbox{SU}(q)}\;.   
\end{equation}
If a subset of the Dynkin labels vanish, for instance, 
$$
a_{p+m}=a_{p+m+1}= \ldots =a_{N-q-n}= 0\;, \qquad p+q+m+n\leq N\;, 
$$
the coset (\ref{5.22}) is further restricted to 
\begin{equation}\label{5.23}
{\mbox{SU}(N)\over [\mbox{U}(1)]^{m+n}\times \mbox{SU}(p)\times 
\mbox{SU}(q)\times \mbox{SU}(N-p-q-m-n+2)}\;.   
\end{equation} 

\subsubsection{$(0,q)$ BPS states} 

Superfields which do not depend on the last  $q$ $\bar\theta$'s 
(or, alternatively, on the first $p$ $\theta$'s) are obtained 
by multiplying G-analytic objects by left- (or right-) handed 
chiral ones: 
$$
{q\over 2N}\mbox{ BPS:}\qquad 
W^{[a_1,a_{2},\ldots,a_{N-q},0,\ldots,0]}_{\alpha_1\ldots\alpha_{2j_1}} 
(\theta_{1},\ldots,\theta_{N}, 
\bar\theta^1,\ldots,\bar\theta^{N-q}) 
$$ 
\begin{equation}\label{5.24}
=w_{\alpha_1\ldots\alpha_{2j_1}}\;\Phi^s\; 
(W^{1})^{a_1}(W^{12})^{a_{2}}\ldots (W^{12\ldots N-q})^{a_{N-q}}   
\end{equation}
where $s\geq0$ is an integer and 
\begin{equation}\label{5.25}
  1\leq q \leq N-1\;.
\end{equation}
Note that the fraction of supersymmetry preserved by a $(0,q)$ BPS 
state ranges as follows: 
\begin{equation}\label{5.25'}
{1\over 2N} \leq {q\over 2N} \leq {N-1\over 2N}\;.
\end{equation}

Such states have the last $q-1$ $\mbox{SU}(N)$ Dynkin labels 
vanishing. The remaining quantum numbers are: 
\begin{equation}\label{5.26}
  \ell = 1+j_1+s+\sum_{k=p}^{N-q}a_k\;, \quad j_2=0\;, 
\quad r = -1-j_1-s+\sum_{k=p}^{N-q}({2k\over 
N}-1)a_k\;. 
\end{equation}
Generically, such superfields live in the harmonic space 
\begin{equation}\label{5.27}
{\mbox{SU}(N)\over [\mbox{U}(1)]^{N-q}\times  \mbox{SU}(q)}\;.   
\end{equation}
If a subset of the Dynkin labels vanish, for instance,
$$
a_{i}= 0\;, \qquad 1 \leq n\leq N-q-1\;, 
$$
the coset (\ref{5.22}) is further restricted to 
\begin{equation}\label{5.28}
{\mbox{SU}(N)\over [\mbox{U}(1)]^{N-q-n}\times  \mbox{SU}(q)\times 
\mbox{SU}(n+1)}\;.   
\end{equation} 

\subsubsection{Chiral BPS states} 

These are described by superfields which do not depend on all of 
the $\bar\theta$'s (or, alternatively, on the $\theta$'s), i.e. 
which are left- (or right-) handed chiral: 
\begin{equation}\label{5.29}
{1\over 2}\mbox{ BPS:}\qquad W_{\alpha_1\ldots\alpha_{2j_1}} 
(\theta_{1},\ldots,\theta_{N}) = 
w_{\alpha_1\ldots\alpha_{2j_1}}\;\Phi^s\;.  
\end{equation}
They are $\mbox{SU}(N)$ singlets. The remaining quantum numbers 
are: 
\begin{equation}\label{5.30}
  \ell = 1+j_1+s\;, \quad j_2=0\;, \quad r = -1-j_1-s\;. 
\end{equation}
The chiral superfields are harmonic-independent.

\setcounter{equation}0 
\section{The six-dimensional case}
 
The method described above can also be applied to the 
superconformal algebras $\mbox{OSp}(8^*/2N)$ in six dimensions. We 
will first examine the consequences of G-analyticity and conformal 
supersymmetry and find out the relation to BPS states. Then we 
will construct UIR's of $\mbox{OSp}(8^*/2N)$ by multiplying 
supersingletons. The results exactly match the general 
classification of UIR's of $\mbox{OSp}(8^*/2N)$ of Ref. 
\cite{Minw2}. Some of the results relevant to the cases $N=1,2$  
have already been presented in \cite{FS3}.  

\subsection{The conformal superalgebra $\mbox{OSp}(8^*/2N)$ 
and Grassmann analyticity}

The part of the conformal superalgebra $\mbox{OSp}(8^*/2N)$
relevant to our discussion is given below:
\begin{eqnarray}
  &&\{Q^i_\alpha, Q^j_\beta\} = 2\Omega^{ij}\gamma^\mu_{\alpha\beta}P_\mu
\;, \label{6.1}\\
  &&\{S^{\alpha\; i}, S^{\beta\; j}\} = 2\Omega^{ij}\gamma_\mu^{\alpha\beta}K^\mu
\;, \label{6.1'}\\
  &&\{Q^i_\alpha, S^{\beta\; j}\} = i\Omega^{ij}(\gamma^{\mu\nu})_\alpha{}^\beta
M_{\mu\nu} + 2\delta^\beta_\alpha (4T^{ij}-i\Omega^{ij}D)\;,
\label{6.2}\\
  &&[D,Q^i_\alpha]= {i\over 2}Q^i_\alpha\;, \qquad [D,S^{\alpha\; i}]= -{i\over 2}S^{\alpha\; i}\;, \label{6.2'}\\
  && [T^{ij},Q^k_\alpha]= -{1\over 2}(\Omega^{ki}Q^{j}_\alpha +
\Omega^{kj}Q^{i}_\alpha)\;, \label{6.3}\\
  && [T^{ij},T^{kl}]= {1\over 2}(\Omega^{ik}T^{lj} + \Omega^{il}T^{kj} +
\Omega^{jk}T^{li} + \Omega^{jl}T^{ki}) \;.\label{6.4}
\end{eqnarray}
Here $Q^i_\alpha$ are the generators of Poincar\'{e} supersymmetry
carrying a right-handed chiral spinor index $\alpha=1,2,3,4$ of
the Lorentz group $\mbox{SU}^*(4)\sim \mbox{SO}(5,1)$ (generators
$M_{\mu\nu}$) and an index $i=1,2,\ldots,2N$ of the fundamental
representation of the R symmetry group $\mbox{USp}(2N)$
(generators $T^{ij}=T^{ji}$); $S^{\beta\; j}$ are the generators
of conformal supersymmetry carrying a left-handed chiral spinor
index; $D$  is the generator of dilations, $P_\mu$ of
translations and $K_\mu$ of conformal boosts. It is convenient to make the non-standard choice of
the symplectic matrix $\Omega^{ij}=-\Omega^{ji}$ with
non-vanishing entries $\Omega^{1\; 2N}=\Omega^{2\; 2N-1}=\ldots
=\Omega^{N\; N+1}=1$. The chiral spinors satisfy a pseudo-reality
condition of the type $\overline{Q_\alpha^i} =
\Omega^{ij}Q_j^\beta c_{\beta\alpha}$ where $c$ is a $4\times 4$
unitary ``charge conjugation" matrix. Note that the generators $M,P,K,D$ form the Lie algebra of $\mbox{SO}(8^*)\sim \mbox{SO}(2,6)$ and the generators $Q,S$ form an $\mbox{SO}(8^*)$ chiral spinor.
 
The standard realization of this superalgebra makes use of the 
superspace 
\begin{equation}
{\mathbb R}^{6\vert 8N} = {\mbox{OSp}(8^*/2N)\over \{K,S,M,D,T\}} 
= (x^\mu,\theta^{\alpha\; i}) \label{6.5} 
\end{equation}
where $\theta^{\alpha\; i}$ is a left-handed spinor. Unlike the 
four-dimensional case,  here chirality is not an option but is 
already built in. The only way to obtain smaller superspaces is 
through Grassmann analyticity. We begin by imposing a single 
condition of G-analyticity (cf. eq. (\ref{18})): 
\begin{equation}\label{6.6}
  q^1_\alpha\Phi(x,\theta)=0
\end{equation}
which amounts to considering the coset
\begin{equation}
{\mathbb A}^{6\vert 4(2N-1)} = {\mbox{OSp}(8^*/2N)\over 
\{K,S,M,D,T,Q^1\}} = (x^\mu,\theta^{\alpha\; 1,2,\ldots,2N-1}) 
\label{6.7} 
\end{equation}
(note that with our conventions $\theta^{\alpha\; 
1}=\theta^\alpha_{2N}, \ldots, \theta^{\alpha\; 
N}=\theta^\alpha_{N+1}$, $\theta^{\alpha\; 
N+1}=-\theta^\alpha_N,\ldots,\theta^{\alpha\; 
2N}=-\theta^\alpha_1$). From the algebra (\ref{6.1})-(\ref{6.4}) 
we obtain 
\begin{eqnarray}
  &&m_{\mu\nu}=0\;, \label{6.8}\\
  &&t^{11}=t^{12}=\ldots=t^{1\; 2N-1}=0\;, \label{6.9}\\
  &&4t^{1\; 2N}+\ell=0\;.\label{6.10}
\end{eqnarray}
Eq. (\ref{6.8}) implies that the superfield $\Phi$ must be a 
Lorentz scalar. In order to interpret eqs. (\ref{6.9}), 
(\ref{6.10}), we need to split the generators of $\mbox{USp}(2N)$ 
into raising operators (corresponding to the positive roots):
\begin{equation}\label{6.111}
  T^{k\; 2N-l}\;, \ \ k=1,\ldots, N,\ l=k,\ldots,2N-k\quad (\mbox{simple if 
$l=k$})\;;
\end{equation}
$[\mbox{U}(1)]^N$ charges: 
\begin{equation}\label{6.112}
  H_k = -2 T^{k\; 2N-k+1}\;, \quad k=1,\ldots, N\;; 
\end{equation}
the remaining generators are lowering operators (corresponding to 
the negative roots). The  Dynkin labels $a_k$ of a  
$\mbox{USp}(2N)$ irrep are defined as follows:
\begin{equation}\label{6.113}
  a_k=H_k-H_{k+1}\;, \ \ k=1,\ldots,N-1\;, \quad a_N=H_N\;,
\end{equation}
so that, for instance, the generator $Q^1$ is the HWS of the 
fundamental irrep $(1,0,\ldots,0)$. 

Now it becomes clear that (\ref{6.9}) is part of the 
$\mbox{USp}(2N)$ irreducibility conditions whereas (\ref{6.10}) 
relates the conformal dimension to the sum of the Dynkin labels: 
\begin{equation}\label{6.12}
  \ell = 2\sum_{k=1}^N a_k\;.
\end{equation}
Let us denote the highest-weight UIR's of the $\mbox{OSp}(8^*/2N)$ 
algebra by 
$$
{\cal D}(\ell;J_1,J_2,J_3;a_1,\ldots,a_N) 
$$
where $\ell$ is the conformal dimension, $J_1,J_2,J_3$ are the 
$\mbox{SU}^*(4)$ Dynkin labels and $a_k$ are the $\mbox{USp}(2N)$  
Dynkin labels of the first component. Then the G-analytic 
superfields defined above are of the type 
\begin{equation}\label{6.13}
 \Phi(\theta^{1,2,\ldots,2N-1}) \ \Leftrightarrow \ 
{\cal D}(2\sum_{k=1}^N a_k;0,0,0;a_1,\ldots,a_N)\;. 
\end{equation}

The next step is to add the generator $Q^2_\alpha$ to the 
superspace coset denominator:
\begin{equation}
{\mathbb A}^{6\vert 4(2N-2)} = {\mbox{OSp}(8^*/2N)\over 
\{K,S,M,D,T,Q^1,Q^2\}} = (x^\mu,\theta^{\alpha\; 
1,2,\ldots,2N-2})\;. \label{6.14} 
\end{equation}  
This implies the new constraints
\begin{eqnarray}
  &&4t^{2\; 2N-1}+\ell=0\quad \Rightarrow \  a_1=0\;,  \label{6.15}\\
  && t^{2\; 2N}=0\;. \label{6.16}
\end{eqnarray}
Note that the vanishing of the lowering operator $t^{2\; 2N}$ 
means that the subalgebra $\mbox{SU}(2)\subset \mbox{USp}(2N)$ 
formed by $t^{1\; 2N-1}$, $t^{2\; 2N}$ and $t^{1\; 2N}-t^{2\; 
2N-1}$ acts trivially on the particular $\mbox{USp}(2N)$ irreps. 
This is equivalent to setting $a_1=0$, as in (\ref{6.15}). Thus, 
the new G-analytic superfields are of the type 
\begin{equation}\label{6.17}
 \Phi(\theta^{1,2,\ldots,2N-2}) \ \Leftrightarrow \ 
{\cal D}(2\sum_{k=2}^N a_k;0,0,0;0,a_2,\ldots,a_N)\;. 
\end{equation}

From (\ref{6.1}) it is clear that we can go on in the same manner 
until we remove half of the $\theta$'s, namely 
$\theta^{N+1},\ldots,\theta^{2N}$. Each time we have to set a new 
Dynkin label to zero. We can summarize by saying that the 
superconformal algebra $\mbox{OSp}(8^*/2N)$ admits the following 
short UIR's corresponding to BPS states: 
\begin{equation}\label{6.18}
  {p\over 2N}\mbox{ BPS}:\quad 
{\cal D}(2\sum_{k=p}^N a_k;0,0,0;0,\ldots,0,a_p,\ldots,a_N)\;, 
\quad p=1,\ldots,N\;. 
\end{equation}

\subsection{Supersingletons}

There exist three types of massless multiplets in six dimensions 
corresponding to ultrashort UIR's (supersingletons) of 
$\mbox{OSp}(8^*/2N)$ (see, e.g., \cite{GT} for the case $N=2$). 
All of them can be formulated in terms of constrained superfields 
as follows.

{\sl (i)} The first type is described by a superfield 
$W^{\{i_1\ldots i_n\}}(x,\theta)$, $1\leq n \leq N$, which is 
antisymmetric and traceless in the external $\mbox{USp}(2N)$ 
indices (for even $n$ one can impose a reality condition). It 
satisfies the constraint (see \cite{HSiT} and \cite{Park}) 
\begin{equation}\label{6.19}
  D^{(k}_\alpha W^{\{i_1)i_2\ldots i_n\}}=0 \qquad \Rightarrow \ {\cal 
D}(2;0,0,0;0,\ldots,0,a_{n}=1,0,\ldots,0) 
\end{equation}
where the spinor covariant derivatives obey the supersymmetry 
algebra 
\begin{equation}\label{6.21}
  \{ D^i_\alpha, D^j_\beta\} = 
-2i\Omega^{ij}\gamma^\mu_{\alpha\beta}\partial_\mu\;. 
\end{equation}
The components of this superfield are massless fields. In the case 
$N=n=1$ this is the on-shell $(1,0)$ hypermultiplet and for 
$N=n=2$ it is the on-shell $(2,0)$ tensor multiplet 
\cite{HSiT,bsvp}.  

{\sl (ii)} The second type is described by a (real) superfield 
without external indices, $w(x,\theta)$ obeying the constraint 
\begin{equation}\label{6.23}
 D^{(i}_{[\alpha} D^{j)}_{\beta]} w = 0 \qquad \Rightarrow \ {\cal 
D}(2;0,0,0;0,\ldots,0)\;.
\end{equation}

{\sl (iii)} Finally, there exists an infinite series of 
multiplets described by superfields with $n$ totally symmetrized 
external Lorentz spinor indices, 
$w_{(\alpha_1\ldots\alpha_n)}(x,\theta)$ (they can be made real 
in the case of even $n$) obeying the constraint
\begin{equation}\label{6.24}
  D^i_{[\beta}w_{(\alpha_1]\ldots\alpha_n)} = 0 \qquad \Rightarrow \ {\cal 
D}(2+n/2;n,0,0;0,\ldots,0)\;. 
\end{equation}

As shown in ref. \cite{FS3}, the six-dimensional massless 
conformal fields only carry reps $(J_1,0)$ of the little group 
$\mbox{SU}(2)\times \mbox{SU}(2)$ of a light-like particle 
momentum. This result is related to the analysis of conformal 
fields in $d$ dimensions \cite{Siegel1,AL}. This fact implies 
that massless superconformal multiplets are classified by a 
single $\mbox{SU}(2)$ and $\mbox{USp}(2N)$ R-symmetry and are 
therefore identical to massless super-Poincar\'e multiplets in 
five dimensions. Some physical implication of the above 
circumstance have recently been discussed in ref. \cite{HULL2} 
where it was suggested that certain strongly coupled $d=5$ 
theories effectively become six-dimensional.

\subsection{Harmonic superspace}

The massless multiplets {\sl (i), (ii)} admit an alternative 
formulation in harmonic superspace (see \cite{HStT}-\cite{Howe} 
for $N=1,2$). The advantage of this formulation is that the 
constraints (\ref{6.19}) become conditions for G-analyticity. We 
introduce harmonic variables describing the coset 
$\mbox{USp}(2N)/[\mbox{U}(1)]^N$: 
\begin{equation}\label{6.25}
  u\in \mbox{USp}(2N): \qquad u^I_iu^i_J = \delta^I_J\;, 
\ \ u^I_i \Omega^{ij}u^J_j = \Omega^{IJ}\;, \ \  u^I_i= 
(u^i_I)^*\;. 
\end{equation}
Here the indices $i,j$ belong to the fundamental representation of  
$\mbox{USp}(2N)$ and $I,J$ are labels corresponding to the 
$[\mbox{U}(1)]^N$ projections. The harmonic derivatives  
\begin{equation}\label{6.26}
  D^{IJ} = \Omega^{K(I}u^{J)}_i{\partial\over\partial u^K_i}
\end{equation}
form the algebra of $\mbox{USp}(2N)_R$ (see (\ref{6.4})) realized 
on the indices $I,J$ of the harmonics. 

Let us now project the defining constraint (\ref{6.19}) with the 
harmonics $u^K_k u^1_{i_1}\ldots u^n_{i_n}$, $K=1,\ldots,n$: 
\begin{equation}\label{6.27}
D^1_\alpha W^{12\ldots n} = D^2_\alpha W^{12\ldots n}= \ldots =  
D^n_\alpha W^{12\ldots n} =0 
\end{equation}
where $D^{K}_\alpha = D^i_\alpha u^{K}_i$ and $W^{12\ldots 
n}=W^{\{i_1\ldots i_n\}}u^1_{i_1}\ldots u^n_{i_n}$. Indeed, the 
constraint (\ref{6.19}) now takes the form of a G-analyticity 
condition. In the appropriate basis in superspace the solution to 
(\ref{6.27}) is a short superfield depending on part of the odd 
coordinates: 
\begin{equation}\label{6.28}
W^{12\ldots n}(x_A,\theta^1,\theta^2,\ldots, \theta^{2N-n},u)\;. 
\end{equation}
In addition to (\ref{6.27}), the projected superfield $W^{12\ldots 
n}$ automatically satisfies the $\mbox{USp}(2N)$ harmonic 
irreducibility conditions 
\begin{equation}\label{6.29}
   D^{K\; 2N-K}W^{12} = 0\;, \quad K=1,\ldots,N
\end{equation}
(only the simple roots of $\mbox{USp}(2N)$ are shown). The 
equivalence between the two forms of the constraint follows from 
the obvious properties of the harmonic products $u^K_{[k} u^K_{i]} 
=0$ and $\Omega^{ij}u^K_iu^L_j=0$ for $1\leq K < L\leq n$. The 
harmonic constraints (\ref{6.29}) make the superfield ultrashort. 

Finally, in case (ii), projecting the constraint (\ref{6.23}) with 
$u^I_iu^I_j$ where $I=1,\ldots,N$ (no summation), we obtain the 
condition 
\begin{equation}\label{6.30'}
  D^I_\alpha D^I_\beta w=0\;.
\end{equation}
It implies that the superfield $w$ is {\sl linear} in each 
projection $\theta^{\alpha I}$.

\subsection{Series of UIR's of $\mbox{OSp}(8^*/2N)$ and shortening}
\label{short6} 

It is now clear that we can realize the BPS series of UIR's 
(\ref{6.18}) as products of the different G-analytic superfields 
(supersingletons) (\ref{6.27}).\footnote{As a bonus, we also 
prove the unitarity of these series, since they are obtained by 
multiplying massless unitary multiplets.} BPS shortening is 
obtained by setting the first $p-1$ $\mbox{USp}(2N)$ Dynkin 
labels to zero: 
\begin{equation}\label{6.34}
{p\over 2N}\ \mbox{BPS}:\ \ W^{[0,\ldots,0,a_p,\ldots,a_N]} 
(\theta^1,\theta^2,\ldots,\theta^{2N-p}) =  (W^{1\ldots 
p})^{a_p}\ldots (W^{1\ldots N})^{a_N}  
\end{equation}
(note that even if $a_1\neq 0$ we still have $1/2N$ shortening).

We remark that our harmonic coset $\mbox{USp}(2N)/[\mbox{U}(1)]^N$ 
is effectively reduced to
\begin{equation}\label{6.35}
  {\mbox{USp}(2N)\over \mbox{U}(p)\times [\mbox{U}(1)]^{N-p}}
\end{equation}
in the case of $p/2N$ BPS shortening (just as it happened in four 
dimensions).  Such a smaller harmonic space was used in  Ref. 
\cite{Howe} to formulate the $(2,0)$ tensor multiplet. 

A study of the most general UIR's of $\mbox{OSp}(8^*/2N)$ 
(similar to the one of Ref. \cite{dp} for the case of 
$\mbox{SU}(2,2/N)$) is presented in Ref. \cite{Minw2}. We can 
construct these UIR's by multiplying the three types of 
supersingletons above:
\begin{equation}\label{6.32}
  w_{\alpha_1\ldots\alpha_{m_1}}w_{\beta_1\ldots\beta_{m_2}}
w_{\gamma_1\ldots\gamma_{m_3}}\; w^k\; W^{[a_1,\ldots,a_N]}
\end{equation}
where $m_1\geq m_2 \geq m_3$ and the spinor indices are arranged 
so that they form an $\mbox{SU}^*(4)$ UIR with Young tableau 
$(m_1,m_2,m_3)$ or Dynkin labels 
$J_1=m_1-m_2,J_2=m_2-m_3,J_3=m_3$. Thus we obtain four distinct 
series:
\begin{eqnarray}
  \mbox{A)}&& \ell
\geq 6 +{1\over 2}(J_1+2J_2+3J_3)+2\sum_{k=1}^N a_k\;; \nonumber\\
  \mbox{B)}&& J_3=0\;,  \qquad \ell
\geq 4 +{1\over 2}(J_1+2J_2)+2\sum_{k=1}^N a_k\;; \nonumber\\
  \mbox{C)}&& J_2=J_3=0\;, \qquad \ell
\geq 2 +{1\over 2}J_1+2\sum_{k=1}^N a_k\;; \nonumber\\
  \mbox{D)}&& J_1=J_2=J_3= 0\;, \qquad \ell
= 2\sum_{k=1}^N a_k\;. \label{6.33'}
\end{eqnarray}
The superconformal bound is saturated when $k=0$ in (\ref{6.32}). 
Note that the values of the conformal dimension we can obtain are 
``quantized" since the factor $w^k$ has $\ell=2k$ and $k$ must be 
a non-negative integer to ensure unitarity. With this restriction 
eq. (\ref{6.33'}) reproduces the results of Ref. \cite{Minw2}. 
However, we cannot comment on the existence of a ``window" of 
dimensions $2 +{1\over 2}J_1+2\sum_{k=1}^N a_k\leq \ell \leq 4 
+{1\over 2}J_1+2\sum_{k=1}^N a_k$ conjectured in \cite{Minw2}. 
\footnote{In a recent paper \cite{FFr} the UIR's of the 
six-dimensional conformal algebra $\mbox{SO}(2,6)$ have been 
classified. Note that the superconformal bound in case A (with 
all $a_i=0$) is stronger that the purely conformal unitarity 
bounds found in \cite{FFr}.}

In the generic case the multiplet (\ref{6.32}) is ``long", but for 
certain special values of the dimension some shortening can take 
place \cite{Minw2}. We can immediately identify all these short 
multiplets. First of all, case D corresponds to BPS shortening. 
In the other cases let us first set $a_i=0$, i.e. no BPS 
multiplets appear in (\ref{6.32}). Then saturating the bound in 
case A (i.e., setting $k=0$) leads to the shortening condition 
(see (\ref{6.24})):
\begin{equation}\label{6.3311}
  \epsilon^{\delta\alpha\beta\gamma}
D^i_\delta(w_{\alpha\ldots\alpha_{m_1}}w_{\beta\ldots\beta_{m_2}} 
w_{\gamma\ldots\gamma_{m_3}}) = 0\ \rightarrow \ \ell = 6 
+{1\over 2}(J_1+2J_2+3J_3)\;.
\end{equation}
Next, in case B we have two possibilities: either we saturate the 
bound ($k=0$) or we use just one factor $w$ ($k=1$). Using 
(\ref{6.23}) and (\ref{6.24}), we find
\begin{equation}\label{6.3312}
  \epsilon^{\delta\gamma\alpha\beta} D^i_\gamma(w_{\alpha\ldots\alpha_{m_1}}
w_{\beta\ldots\beta_{m_2}}) = 0\ \rightarrow \ \ell = 4 +{1\over 
2}(J_1+2J_2)\;;
\end{equation}
\begin{equation}\label{6.3313}
 \epsilon^{\delta\gamma\alpha\beta}D^{(i}_\delta
D^{j)}_\gamma(w\;w_{\alpha\ldots\alpha_{m_1}} 
w_{\beta\ldots\beta_{m_2}}) = 0\ \rightarrow \ \ell = 6 +{1\over 
2}(J_1+2J_2)\;.
\end{equation}
Similarly, in case C with $J_1\neq0$ we have three options, namely 
setting $k=0\ \rightarrow \ \ell = 2+{1\over 2}J_1$ (which 
corresponds to the supersingleton defining constraint 
(\ref{6.24})) or $k=1,2$ which gives:
\begin{equation}\label{6.3314}
 \epsilon^{\delta\gamma\beta\alpha}D^{(i}_\gamma
D^{j)}_\beta(w\;w_{\alpha\ldots\alpha_{m_1}}) = 0\ \rightarrow \ 
\ell = 4 +{1\over 2}J_1\;,
\end{equation}
\begin{equation}\label{6.3315}
 \epsilon^{\delta\gamma\beta\alpha}D^{(i}_\delta D^j_\gamma
D^{k)}_\beta(w^2\;w_{\alpha\ldots\alpha_{m_1}}) = 0\ \rightarrow 
\ \ell = 6 +{1\over 2}J_1\;.
\end{equation}
Finally, in case C with $J_1=0$ we can take the scalar 
supersingleton (\ref{6.23}) itself, i.e. set $k=1\ \rightarrow \ 
\ell = 2$, or set $k=2,3$:
\begin{equation}\label{6.3316}
  \epsilon^{\delta\gamma\beta\alpha}D^{(i}_\gamma D^j_\beta
D^{k)}_\alpha(w^2) = 0\ \rightarrow \ \ell = 4\;,
\end{equation}
\begin{equation}\label{6.3317}
  \epsilon^{\delta\gamma\beta\alpha}D^{(i}_\delta D^j_\gamma
D^k_\beta D^{l)}_\alpha(w^3) = 0\ \rightarrow \ \ell = 6\;.
\end{equation}

Introducing $\mbox{USp}(2N)$ quantum numbers into the above 
shortening conditions is achieved by multiplying the short 
multiplets by a BPS object. The new short multiplets satisfy the 
corresponding $\mbox{USp}(2N)$ projections of eqs. (\ref{6.23}), 
(\ref{6.24}), (\ref{6.3311})-(\ref{6.3317}). We call such objects 
``intermediate short".

\setcounter{equation}0 
\section{The three-dimensional case}

In this section we carry out the analysis of the $d=3$ $N=8$ 
superconformal algebra $\mbox{OSp}(8/4,\mathbb{R})$ in a way 
similar to the above. Some of the results have already been 
presented in \cite{FS2}. As in the previous cases, our results 
could easily be extended to $\mbox{OSp}(N/4,\mathbb{R})$ 
superalgebras with arbitrary $N$. The $N=2$ and $N=3$ cases were 
considered in Ref. \cite{Torino}.

\subsection{The conformal superalgebra $\mbox{OSp}(8/4,\mathbb{R})$ 
and Grassmann analyticity}\label{CSGA} 

The part of the conformal superalgebra 
$\mbox{OSp}(8/4,\mathbb{R})$ relevant to our discussion is given 
below: 
\begin{eqnarray}
  && \{Q^i_\alpha, Q^j_\beta\} = 2\delta^{ij} \gamma^\mu_{\alpha\beta} 
P_\mu\;, \label{7.1}\\
  && \{Q^i_\alpha, S^j_\beta\} = \delta^{ij} 
M_{\alpha\beta} + 2\epsilon_{\alpha\beta}  (T^{ij} +  \delta^{ij} 
D) \;, \label{7.2}\\  
  && [T^{ij}, Q^k_\alpha] = i(\delta^{ki} Q^j_\alpha - \delta^{kj} 
Q^i_\alpha)\;,\label{7.3}\\ 
  && [T^{ij}, T^{kl}] = i(\delta^{ik} T^{jl} + \delta^{jl} T^{ik} 
- \delta^{jk} T^{il} - \delta^{il} T^{jk})\;. \label{7.4} 
\end{eqnarray}
Here we find the following generators: $Q^i_\alpha$ of $N=8$ 
Poincar\'{e} supersymmetry carrying a spinor index $\alpha=1,2$ of the 
$d=3$ Lorentz group $\mbox{SL}(2,\mathbb{R})\sim \mbox{SO}(1,2)$ 
(generators $M_{\alpha\beta} = M_{\beta\alpha}$) and a vector 
\footnote{Since $\mbox{SO}(8)$ has three 8-dimensional 
representations, $8_v$, $8_s$ and  $8_c$ related by triality, the 
choice which one to ascribe to the supersymmetry generators is 
purely conventional. In order to be consistent with the other 
$N$-extended $d=3$ supersymmetries where the odd generators always 
belong to the vector representation, we prefer to put an $8_v$ 
index $i$ on the supercharges.} index $i=1,\ldots,8$ of the R 
symmetry group $\mbox{SO}(8)$ (generators $T^{ij}=-T^{ji}$); 
$S^i_\alpha$ of conformal supersymmetry; $P_\mu$, $\mu=0,1,2$, of 
translations; $D$ of dilations. 

The standard realization of this superalgebra makes use of the 
superspace 
\begin{equation}
{\mathbb R}^{3\vert 16} = {\mbox{OSp}(8/4,\mathbb{R})\over 
\{K,S,M,D,T\}} = (x^\mu,\theta^{\alpha\; i})\;. \label{7.5} 
\end{equation}
In order to study G-analyticity we need to decompose the 
generators $Q^i_\alpha$ under $[\mbox{U}(1)]^4\subset 
\mbox{SO}(8)$. Besides the vector representation $8_v$ of 
$\mbox{SO}(8)$ we are also going to use the spinor ones, $8_s$ and 
$8_c$. In this context we find it convenient to introduce the four 
subgroups $\mbox{U}(1)$  by successive reductions: $\mbox{SO}(8)\ 
\rightarrow \ \mbox{SO}(2)\times \mbox{SO}(6)\sim 
\mbox{U}(1)\times \mbox{SU}(4) \ \rightarrow \ 
[\mbox{SO}(2)]^2\times \mbox{SO}(4)\sim [\mbox{U}(1)]^2\times 
\mbox{SU}(2)\times \mbox{SU}(2) \ \rightarrow \ [\mbox{SO}(2)]^4 
\sim [\mbox{U}(1)]^4$. Denoting the four $\mbox{U}(1)$ charges by 
$\pm$, $(\pm)$, $[\pm]$ and $\{\pm\}$, we decompose the three 
8-dimensional representations as follows: 
\begin{eqnarray}
8_v:\quad Q^i &\rightarrow& Q^{\pm\pm}, \ Q^{(\pm\pm)}, \
Q^{[\pm]\{\pm\}},\label{7.10}\\
 8_s:\quad \phi^a &\rightarrow&
\phi^{+(+)[\pm]}, \ \phi^{-(-)[\pm]}, \ \phi^{+(-)\{\pm\}}, \
\phi^{-(+)\{\pm\}}\label{7.11}\\
 8_c:\quad  \sigma^{\dot a} &\rightarrow& \sigma^{+(+)\{\pm\}},
\ \sigma^{-(-)\{\pm\}}, \ \sigma^{+(-)[\pm]}, \ 
\sigma^{-(+)[\pm]}\label{7.12}
\end{eqnarray}
The definition of the charge operators $H_i$, $i=1,2,3,4$ can be 
read off from the corresponding projections of the relation 
(\ref{7.2}):  
\begin{eqnarray}
 \{Q^{++}_\alpha, S^{--}_\beta\} &=& {1\over 2} M_{\alpha\beta} + 
\epsilon_{\alpha\beta}  (D-{1\over 2}H_1) \;, \nonumber\\ 
  \{Q^{(++)}_\alpha, S^{(--)}_\beta\} &=& {1\over 2} M_{\alpha\beta} + 
\epsilon_{\alpha\beta}  (D - {1\over 2}H_2) \;, \nonumber\\ 
   \{Q^{[+]\{+\}}_\alpha, S^{[-]\{-\}}_\beta\} &=& {1\over 2} M_{\alpha\beta} + 
\epsilon_{\alpha\beta}  (D - {1\over 2}H_3  - {1\over 2}H_4) \;, 
\nonumber\\  
   \{Q^{[+]\{-\}}_\alpha, S^{[-]\{+\}}_\beta\} &=& -{1\over 2} M_{\alpha\beta} - 
\epsilon_{\alpha\beta}  (D - {1\over 2}H_3  + {1\over 2}H_4) \; . 
\label{7.9} 
\end{eqnarray}
In this notation we have
\begin{eqnarray}
  &&[H_1, Q^{\pm\pm}_\alpha ] = [H_2, Q^{(\pm\pm)}_\alpha ] = 
\pm 2i Q^{\pm\pm}_\alpha\;, \nonumber\\
  &&[H_3, Q^{[\pm]\{\pm\}}] = [H_4, Q^{[\pm]\{\pm\}}] = \pm i 
Q^{[\pm]\{\pm\}}\;. \label{7.13} 
\end{eqnarray}

Let us denote a quasi primary superconformal field of the 
$\mbox{OSp}(8/4,\mathbb{R})$ algebra by the quantum numbers of its 
HWS: 
\begin{equation}\label{555}
 {\cal D}(\ell; J; a_1,a_2,a_3,a_4)  
\end{equation}
where $\ell$ is the conformal dimension, $J$ is the Lorentz spin 
and $a_i$ are the Dynkin labels (see, e.g., \cite{FSS}) of the 
$\mbox{SO}(8)$ R symmetry. In fact, in our scheme the natural 
labels are the four charges $h_i$ (the eigenvalues of $H_i$). They 
are related to the Dynkin labels as follows: 
\begin{eqnarray}
  &&h_1= 2(a_1+a_2)+a_3+a_4\;, \nonumber\\
  &&h_2= 2a_2+a_3+a_4\;, \label{7.14}\\
  &&h_3=a_3\;,\quad h_4=a_4\;, \nonumber 
\end{eqnarray}
or, inversely,
\begin{equation}\label{7.14'}
  a_1= {1\over 2}(h_1-h_2)\;, \quad a_2= {1\over 2}(h_2-h_3-h_4)\;, \quad 
a_3=h_3\;, \quad a_4=h_4\;.
\end{equation}
A HWS $|a_i\rangle$ of $\mbox{SO}(8)$ is by definition annihilated 
by the positive simple roots of the $\mbox{SO}(8)$ algebra: 
\begin{equation}\label{7.15}
  T^{[++]}|a_i\rangle = T^{\{++\}}|a_i\rangle = 
T^{++(--)}|a_i\rangle = T^{(++)[-]\{-\}}|a_i\rangle = 0\;.
\end{equation}

In order to build G-analytic superspaces we have to add one or 
more projections of $Q^i_\alpha$ to the coset denominator. In 
choosing the subset of projections we have to make sure that: i) 
they anticommute among themselves; ii) the subset is closed under 
the action of the raising operators of $\mbox{SO}(8)$ 
(\ref{7.15}). Then we have to examine the consistency of the 
vanishing of the chosen projections with the conformal 
superalgebra (\ref{7.9}). Thus we find the following sequence of 
G-analytic superspaces corresponding to BPS states: 
\begin{eqnarray}
 {1\over 8}  \mbox{ BPS}: && \left\{
  \begin{array}{l}
    q_\alpha^{++}\Phi=0\ \rightarrow \\ 
    \Phi(\theta^{++},\theta^{(\pm\pm)},\theta^{[\pm]\{\pm\}})\\
    {\cal D}(a_1+a_2 + {1\over 2}(a_3+a_4);0;a_1,a_2,a_3,a_4) 
     \end{array}
 \right.\label{7.16}\\
 {1\over 4}  \mbox{ BPS}: &&\left\{
  \begin{array}{l}
    q_\alpha^{++}\Phi=q_\alpha^{(++)}\Phi=0\ \rightarrow \\ 
    \Phi(\theta^{++},\theta^{(++)},\theta^{[\pm]\{\pm\}})\\
    {\cal D}(a_2 + {1\over 2}(a_3+a_4);0;0,a_2,a_3,a_4)
      \end{array}
 \right. \label{7.17}\\
{3\over 8}  \mbox{ BPS}: &&\left\{ 
  \begin{array}{l}
    q_\alpha^{++}\Phi=q_\alpha^{(++)}\Phi=q_\alpha^{[+]\{+\}}\Phi=0\ \rightarrow \\ 
    \Phi(\theta^{++},\theta^{(++)},\theta^{[+]\{\pm\}},\theta^{[-]\{+\}})\\
    {\cal D}({1\over 2}(a_3+a_4);0;0,0,a_3,a_4)
  \end{array}
 \right. \label{7.18}\\
{1\over 2}  \mbox{ BPS (type I)}: &&\left\{ 
  \begin{array}{l}
    q_\alpha^{++}\Phi=q_\alpha^{(++)}\Phi=q_\alpha^{[+]\{\pm\}}\Phi=0\ \rightarrow \\ 
    \Phi(\theta^{++},\theta^{(++)},\theta^{[+]\{\pm\}})\\
    {\cal D}({1\over 2}a_3;0;0,0,a_3,0)
  \end{array}
 \right. \label{7.19}\\
{1\over 2}  \mbox{ BPS (type II)}: &&\left\{ 
  \begin{array}{l}
    q_\alpha^{++}\Phi=q_\alpha^{(++)}\Phi=q_\alpha^{[\pm]\{+\}}\Phi=0\ \rightarrow \\ 
    \Phi(\theta^{++},\theta^{(++)},\theta^{[\pm]\{+\}})\\
    {\cal D}({1\over 2}a_4;0;0,0,0,a_4)
  \end{array}
 \right. \label{7.20}
\end{eqnarray}
Note the existence of two types of $1/2$ BPS states due to the two 
possible subsets of projections of $q^i$ closed under the raising 
operators of $\mbox{SO}(8)$ (\ref{7.15}). 

We remark that in the cases $1/4$, $3/8$ and $1/2$ the states are
annihilated by some of the lowering operators of $\mbox{SO}(8)$. 
This means that certain subalgebras of $\mbox{SO}(8)$ act 
trivially on them: 
\begin{eqnarray}
  {1\over 4}:\ \mbox{SU}(2) &\leftrightarrow& 
\left\{ T^{++(--)}\;, \ T^{--(++)}\;, \ H_1-H_2 
 \right. \label{7.21}\\
{3\over 8}:\ \mbox{SU}(3) &\leftrightarrow& \left\{ 
\begin{array}{l}
   T^{++(--)}\;, \ T^{--(++)}\;, \ H_1-H_2 \\
   T^{(++)[-]\{-\}}\;, \ T^{(--)[+]\{+\}}\;, \ H_2-H_3-H_4
  \end{array}
 \right. \label{7.22}\\
{1\over 2} :\ \mbox{SU}(4)_I &\leftrightarrow& \left\{ 
\begin{array}{l}
   T^{++(--)}\;, \ T^{--(++)}\;, \ H_1-H_2 \\
   T^{(++)[-]\{-\}}\;, \ T^{(--)[+]\{+\}}\;, \ H_2-H_3-H_4 \\
   T^{\{++\}}\;, \ T^{\{--\}}\;, \ H_4
  \end{array}
 \right. \label{7.23}\\
{1\over 2} :\ \mbox{SU}(4)_{II} &\leftrightarrow&  \left\{ 
\begin{array}{l}
   T^{++(--)}\;, \ T^{--(++)}\;, \ H_1-H_2 \\
   T^{(++)[-]\{-\}}\;, \ T^{(--)[+]\{+\}}\;, \ H_2-H_3-H_4 \\
   T^{[++]}\;, \ T^{[--]}\;, \ H_3
  \end{array}
 \right. \label{7.24}
\end{eqnarray} 
These properties are equivalent to the restrictions on the 
possible values of the $\mbox{SO}(8)$ Dynkin labels in 
(\ref{7.16})-(\ref{7.20}). Note that the existence of two types of 
$1/2$ BPS states can be equivalently explained by the two possible 
ways to embed $\mbox{SU}(4)$ in $\mbox{SO}(8)$, as shown in 
(\ref{7.23}) and (\ref{7.24}). 

\subsection{Supersingletons and harmonic superspace}

The supersingletons are the simplest $\mbox{OSp}(8/4,\mathbb{R})$ 
representations of the type (\ref{7.19}) or (\ref{7.20}) and 
correspond to ${\cal D}(1/2; 0; 0,0,1,0)$ or ${\cal D}(1/2; 0; 
0,0,0,1)$. The existence of two distinct types of $d=3$ $N=8$ 
supersingletons has first been noted in Ref. \cite{GNST}. Each of 
them is just a collection of eight Dirac supermultiplets \cite{Fr} 
made out of ``Di" and ``Rac" singletons \cite{ff2}. 

In order to realize the supersingletons in superspace we note that 
the HWS in the two supermultiplets above has spin 0 and the Dynkin 
labels of the $8_s$ or $8_c$ of $\mbox{SO}(8)$, correspondingly. 
Therefore we take a scalar superfield $\Phi_a(x^\mu, 
\theta^\alpha_i)$ (or $\Sigma_{\dot a}(x^\mu, \theta^\alpha_i)$) 
carrying an external $8_s$ index $a$ (or an $8_c$ index $\dot a$). 
These superfields are subject to the following on-shell 
constraints \footnote{See also \cite{Howe} for the description of 
a supersingleton related to ours by $\mbox{SO}(8)$ triality. 
Superfield representations of other $OSp(N/4)$ superalgebras have 
been considered in \cite{IS,FFre}.}: 
\begin{eqnarray}
  \mbox{type I:}&&D^i_\alpha\Phi_a = {1\over 8}\gamma^i_{a\dot 
b}\tilde\gamma^j_{\dot b c} D^j_\alpha\Phi_c\;; \label{7.25}\\ 
  \mbox{type II:}&& D^i_\alpha\Sigma_{\dot a} = {1\over 
8}\tilde\gamma^i_{\dot a b}\gamma^j_{b\dot c} 
D^j_\alpha\Sigma_{\dot c}\;. \label{7.26} 
\end{eqnarray}
The two multiplets consist of a massless scalar in the $8_s$ 
($8_c$) and spinor in the $8_c$ ($8_s$). 

The harmonic superspace description of these supersingletons can 
be realized by taking the harmonic coset \footnote{A formulation 
of the above multiplet in harmonic superspace has been proposed 
in Ref. \cite{Howe} (see also \cite{ZK} and \cite{HL} for a 
general discussion of three-dimensional harmonic superspaces). 
The harmonic coset used in \cite{Howe} is 
$\mbox{Spin}(8)/\mbox{U}(4)$. Although the supersingleton itself 
does indeed live in this smaller coset (see Section \ref{7.4.4}), 
its residual symmetry $U(4)$ would not allow us to multiply 
different realizations of the supersingleton. For this reason we 
prefer from the very beginning to use the coset (\ref{7.27'}) 
with a minimal residual symmetry.} 
\begin{equation}\label{7.27'}
  {\mbox{SO}(8)\over [\mbox{SO}(2)]^4} \ \sim \ {\mbox{Spin}(8)\over
 [\mbox{U}(1)]^4}\;.
\end{equation}  
Since $\mbox{SO}(8)\sim \mbox{Spin}(8)$  has three inequivalent 
fundamental representations, $8_s,8_c,8_v$, following  \cite{GHS} 
we introduce three sets of harmonic variables: 
\begin{equation}\label{7.27}
  u_a^A\;, \ w^{\dot A}_{\dot a}\;, \ v^I_i
\end{equation}
where $A$, $\dot A$ and $I$ denote the decompositions of an $8_s$, 
$8_c$ and $8_v$ index, correspondingly, into sets of four 
$\mbox{U}(1)$ charges (see (\ref{7.10})-(\ref{7.12})). Each of the 
$8\times 8$ real matrices (\ref{7.27}) belongs to the 
corresponding representation of $\mbox{SO}(8)\sim \mbox{Spin}(8)$. 
This implies that they are orthogonal matrices (this is a 
peculiarity of $\mbox{SO}(8)$ due to triality): 
\begin{equation}\label{7.28}
  u_a^A u_a^B = \delta^{AB}\;, \quad w^{\dot A}_{\dot a} w^{\dot B}_{\dot 
a}  = \delta^{\dot A\dot B}\;, \quad v^I_i v^J_i = \delta^{IJ} \;.
\end{equation}
These matrices supply three copies of the group space, and we only 
need one to parametrize the harmonic coset. The condition which 
identifies the three sets \footnote{Although each of the three 
sets of harmonic variables depends on the same 28 parameters, we 
need at least two sets to be able to reproduce all possible 
representations of $\mbox{SO}(8)$.} of harmonic variables is 
\begin{equation}\label{7.29}
  u_a^A (\gamma^I)_{A\dot A} w^{\dot A}_{\dot a} = v^I_i (\gamma^i)_{a\dot 
a}\;. 
\end{equation}

Further, we introduce harmonic derivatives (the covariant 
derivatives on the coset (\ref{7.27'})): 
\begin{equation}\label{7.30}
  D^{IJ} = u^A_a (\gamma^{IJ})^{AB}{\partial\over\partial u^B_a} + 
w^{\dot A}_{\dot a} (\gamma^{IJ})^{\dot A\dot 
B}{\partial\over\partial w^{\dot B}_{\dot a}} + v^{[I}_i 
{\partial\over\partial v^{J]}_{i}}\;. 
\end{equation}
They respect the algebraic relations  (\ref{7.28}), (\ref{7.29}) 
among the harmonic variables and form the algebra of 
$\mbox{SO}(8)$ realized on the indices $A,\dot A, I$ of the 
harmonics. 

We now use the harmonic variables for projecting the 
supersingleton defining constraints (\ref{7.25}), (\ref{7.26}). 
Using the relation (\ref{7.29}) it is easy to show that the 
projections $\Phi^{+(+)[+]}$ and $\Sigma^{+(+)\{+\}}$ satisfy the 
following G-analyticity constraints: 
\begin{eqnarray}
  &&D^{++}\Phi^{+(+)[+]} = D^{(++)}\Phi^{+(+)[+]}=D^{[+]\{\pm\}} 
\Phi^{+(+)[+]} = 0\;, \label{7.31}\\ 
  &&D^{++}\Sigma^{+(+)\{+\}} = D^{(++)}\Sigma^{+(+)\{+\}}=D^{[+]\{\pm\}} 
\Sigma^{+(+)\{+\}} = 0 \label{7.31'} 
\end{eqnarray}
where $D^I_\alpha = v^I_iD^i_\alpha$, $\Phi^A = u^A_a\Phi_a$ and 
$\Sigma^{\dot A} = w^{\dot A}_{\dot a}\Sigma_{\dot a}$. This is 
the superspace realization of the 1/2 BPS shortening conditions 
(\ref{7.19}), (\ref{7.20}). In the appropriate basis in superspace 
$\Phi^{+(+)[+]}$ and $\Sigma^{+(+)\{+\}}$ depend on different 
halves of the odd variables as well as on the harmonic variables:
\begin{eqnarray}
  \mbox{type I}:&& \Phi^{+(+)[+]}
(x_A,\theta^{++}, \theta^{(++)}, \theta^{[+]\{\pm\}}, u,w) \;, 
\label{7.32}\\ 
 \mbox{type II}: && \Sigma^{+(+)\{+\}}(x_A,\theta^{++},\theta^{(++)},
\theta^{[\pm]\{+\}}, u,w)\;.\label{7.32'} 
\end{eqnarray}

In addition to the G-analyticity constraints (\ref{7.31}), 
(\ref{7.31'}), the on-shell superfields $\Phi^{+(+)[+]}$, 
$\Sigma^{+(+)\{+\}}$ are subject to the $\mbox{SO}(8)$ 
irreducibility harmonic conditions obtained from (\ref{7.15}) by 
replacing the $\mbox{SO}(8)$ generators by the corresponding 
harmonic derivatives. The combination of the latter with eq. 
(\ref{7.31}) is equivalent to the original constraint 
(\ref{7.25}). 

It should be stressed that $\Phi^{+(+)[+]}$, $\Sigma^{+(+)\{+\}}$ 
automatically satisfy additional harmonic constraints involving 
lowering operators of $\mbox{SO}(8)$ (cf. (\ref{7.23}) and 
(\ref{7.24})). As mentioned earlier, this means that the 
supersingleton harmonic superfields effectively live in the 
smaller harmonic coset $\mbox{Spin}(8)/\mbox{U}(4)$. 

\subsection{$\mbox{OSp}(8/4,\mathbb{R})$ supersingleton composites}

One way to obtain short multiplets of $\mbox{OSp}(8/4,\mathbb{R})$ 
is to multiply different analytic superfields describing the type 
I supersingleton. The point is that above we chose a particular 
projection of, e.g., the defining constraint (\ref{7.25}) which 
lead to the analytic superfield  $\Phi^{+(+)[+]}$. In fact, we 
could have done this in a variety of ways, each time obtaining 
superfields depending on different halves of the total number of 
odd variables. Leaving out the $8_v$ lowest weight $\theta^{--}$, 
we can have four distinct but equivalent analytic descriptions of 
the type I supersingleton: 
\begin{eqnarray}
  &&\Phi^{+(+)[+]}
(\theta^{++}, \theta^{(++)}, \theta^{[+]\{+\}}, 
\theta^{[+]\{-\}})\;, \nonumber\\ 
  &&\Phi^{+(+)[-]} (\theta^{++}, \theta^{(++)}, 
\theta^{[-]\{+\}}, \theta^{[-]\{-\}})\;, \nonumber\\ 
  &&\Phi^{+(-)\{+\}} (\theta^{++}, 
\theta^{(--)}, \theta^{[+]\{+\}}, \theta^{[-]\{+\}})\;, 
\nonumber\\ 
  &&\Phi^{+(-)\{-\}} 
(\theta^{++}, \theta^{(--)}, \theta^{[+]\{-\}}, 
\theta^{[-]\{-\}})\;. \label{7.34} 
\end{eqnarray}
Then we can multiply them in the following way: 
\begin{equation}\label{7.35}
  (\Phi^{+(+)[+]})^{p+q+r+s}(\Phi^{+(+)[-]})^{q+r+s} 
(\Phi^{+(-)\{+\}})^{r+s}(\Phi^{+(-)\{-\}})^{s} 
\end{equation}
thus obtaining three series of $\mbox{OSp}(8/4,\mathbb{R})$ UIR's 
exhibiting $1/8$, $1/4$ or $1/2$ BPS shortening: 
\begin{eqnarray}
 {1\over 8}  \mbox{ BPS:} && {\cal D}(a_1+a_2 + {1\over 
2}(a_3+a_4), 0; a_1,a_2,a_3,a_4)\;, \quad a_1-a_4 = 2s \geq 0\;;    
 \nonumber\\
 {1\over 4}  \mbox{ BPS:} && {\cal D}(a_2 + {1\over 2}a_3, 0; 
0,a_2,a_3,0)\;;  \label{7.36}\\ 
 {1\over 2}  \mbox{ BPS:} && {\cal D}({1\over 2}a_3, 0; 0,0,a_3,0)  
\nonumber \end{eqnarray} 
where
\begin{equation}\label{7.37'}
a_1=r+2s\;, \quad a_2= q\;, \quad a_3=p\;, \quad a_4=r\;.
\end{equation}

We see that multiplying only one type of supersingletons cannot 
reproduce the general result of Section \ref{CSGA} for all 
possible short multiplets. Most notably, in (\ref{7.36}) there is 
no $3/8$ series. The latter can be obtained  by mixing the two 
types of supersingletons: 
\begin{equation}\label{7.37}
[\Phi^{+(+)[+]}(\theta^{++},\theta^{(++)},\theta^{[+]\{\pm\}})]^{a_3} 
[\Sigma^{+(+)\{+\}}(\theta^{++},\theta^{(++)},\theta^{[\pm]\{+\}})]^{a_4}
\end{equation}
(or the same with $\Phi$ and $\Sigma$ exchanged). Counting the 
charges and the dimension, we find exact matching with the series 
(\ref{7.18}):
\begin{equation}\label{7.37''}
 {3\over 8}  \mbox{ BPS:} \quad {\cal D}({1\over 2}(a_3+a_4);0;0,0,a_3,a_4)\;.  
\end{equation}
Further, mixing two realizations of type I and one of type II 
supersingletons, we can construct the 1/4 series 
\begin{equation}\label{7.38}
   [\Phi^{+(+)[+]}]^{a_2+a_3}[\Phi^{+(+)[-]}]^{a_2}
[\Sigma^{+(+)\{+\}}]^{a_4} 
\end{equation}
which corresponds to (\ref{7.17}):
\begin{equation}\label{7.37'''}
 {1\over 4}  \mbox{ BPS:} \quad 
{\cal D}(a_2 + {1\over 2}(a_3+a_4);0;0,a_2,a_3,a_4)\;.  
\end{equation} Finally, the full 1/8 series 
(\ref{7.16}) (i.e., without the restriction $a_1-a_4 = 2s\geq 0$ 
in (\ref{7.36})) can be obtained in a variety of ways. 

In this section we have analyzed all short highest-weight UIR's  
of the $\mbox{OSp}(8/4,\mathbb{R})$ superalgebra whose HWS's are 
annihilated by part of the super-Poincar\'{e} odd generators. The 
number of distinct possibilities have been shown to correspond to 
different BPS conditions on the HWS. When the algebra is 
interpreted on the $AdS_4$ bulk, for which the 3d superconformal 
field theory corresponds to the boundary M-2 brane dynamics, 
these states appear as BPS massive excitations, such as K-K 
states or AdS black holes, of M-theory on $AdS_4\times S^7$. 
Since in M-theory there is only one type of supersingleton 
related to the M-2 brane transverse coordinates \cite{Duff1}, 
according to our analysis massive states cannot be 3/8 BPS 
saturated, exactly as it happens in M-theory on $M^4\times T^7$. 
Indeed, the missing solution was also noticed in Ref. 
\cite{Duff2} by studying $AdS_4$ black holes in gauged $N=8$ 
supergravity. Curiously, in the ungauged theory, which is in some 
sense the flat limit of the former, the 3/8 BPS states are 
forbidden \cite{FMG} by the underlying $E_{7(7)}$ symmetry of 
$N=8$ supergravity \cite{CJ}.

\subsection{Series of UIR's of $\mbox{OSp}(8/4,\mathbb{R})$}\label{short3}

In the cases of even dimension $d=4,6$ we had supersingleton 
superfields carrying either $R$ symmetry indices or Lorentz 
indices or just conformal dimension. Multiplying them we were 
able to reproduce the corresponding general series of UIR's. In 
the case $d=3$ the situation is different, since we only have two 
supersingletons carrying $\mbox{SO}(8)$ spinor indices. 
Multiplying them we could construct the short objects of the BPS 
type considered above. Yet, for reproducing the most general 
UIR's (see \cite{Minw2}), we need short objects with spin but 
without $\mbox{SO}(8)$ indices. These arise in the form of 
conserved currents. The simplest one is a Lorentz scalar and an 
$\mbox{SO}(8)$ singlet $w$ of dimension $\ell=1$. It can be 
realized as a bilinear of two supersingletons of the same type, 
e.g., $w=\Phi_a \Phi_a$ or $w=\Sigma_{\dot a}\Sigma_{\dot a}$. 
Using (\ref{7.25}) or (\ref{7.26}) one can show that it satisfies 
the constraint (a non-BPS shortness condition) 
\begin{equation}\label{7.001}
  D^i_\alpha D^{j\alpha} \; w = 
{1\over 8}\delta^{ij} D^k_\alpha D^{k\alpha} \; w \;.
\end{equation}
The other currents carry $\mbox{SL}(2,\mathbb{R})$ spinor indices, 
$w_{\alpha_1\ldots\alpha_{2J}}$, have dimension $\ell=1 + J$ and 
satisfy the constraint \cite{Park3d} 
\begin{equation}\label{7.002}
  D^{i\alpha}w_{\alpha\alpha_2\ldots\alpha_{2J}} = 0\;.
\end{equation}
They can be constructed as bilinears of the two types of 
supersingletons (for half-integer spin) or of  two copies of the 
same type (for integer spin). For example, the two lowest ones 
($J=1/2$ and $J=1$) are 
\begin{equation}\label{7.0020}
  w_\alpha = \gamma^i_{b\dot b}\left(D^i_\alpha \Phi_b\;\Sigma_{\dot b} - 
\Phi_b\;  D^i_\alpha\Sigma_{\dot b} \right)\;, 
\end{equation}
\begin{equation}\label{7.0021}
  w_{\alpha\beta} =  D^i_{(\alpha}\Phi_a  (\gamma^i\gamma^j)_{ab}
D^j_{\beta)}\Phi'_b + 32i(\Phi_a\; \partial_{\alpha\beta}\Phi'_a 
- \partial_{\alpha\beta}\Phi_a\;  \Phi'_a)\;. 
\end{equation}
They are easily generalized to
\begin{eqnarray}
  &&\quad w_{\alpha_1\ldots\alpha_{2n+1}}= \gamma^i_{b\dot b} \sum_{k=0}^{n}(-1)^k 
\label{7.0022}\\
&&\left(\partial_{(\alpha_1\alpha_2} \ldots 
\partial_{\alpha_{2k-1}\alpha_{2k}} D^i_{\alpha_{2k+1}} \Phi_b \;
\partial_{\alpha_{2k+2}\alpha_{2k+3}} \ldots 
\partial_{\alpha_{2n-1}\alpha_{2n})}\Sigma_{\dot b}\right. \nonumber\\ 
 &&\left. -\partial_{(\alpha_1\alpha_2} \ldots 
\partial_{\alpha_{2k-1}\alpha_{2k}}  \Phi_b\; 
\partial_{\alpha_{2k+1}\alpha_{2k+2}} \ldots 
\partial_{\alpha_{2n-1}\alpha_{2n}} D^i_{\alpha_{2n+1})}\Sigma_{\dot b}\right)\;; 
\nonumber\\ \nonumber\\ 
 &&\quad w_{\alpha_1\ldots\alpha_{2n}}= \sum_{k=0}^{n}(-1)^k 
\label{7.0023}\\
&&\left[ 
\partial_{(\alpha_1\alpha_2} \ldots 
\partial_{\alpha_{2k-1}\alpha_{2k}} D^i_{\alpha_{2k+1}}\Phi_a  (\gamma^i\gamma^j)_{ab} 
D^j_{\alpha_{2k+2}}\partial_{\alpha_{2k+3}\alpha_{2k+4}} \ldots 
\partial_{\alpha_{2n-1}\alpha_{2n})}\Phi'_b\right. \nonumber\\ 
 &&\left. + 32i  \partial_{(\alpha_1\alpha_2} \ldots 
\partial_{\alpha_{2k-1}\alpha_{2k}}  \Phi_a  \;
\partial_{\alpha_{2k+1}\alpha_{2k+2}} \ldots 
\partial_{\alpha_{2n-1}\alpha_{2n})} \Phi'_a\right] \nonumber
\end{eqnarray}
(note that if $n=2m$ the two supersingletons $\Phi_a$ and 
$\Phi'_a$ can be identical).

The generic ``long" UIR of $\mbox{OSp}(8/4,\mathbb{R})$ can now be 
obtained as a product of all of the above short objects: 
\begin{equation}\label{7.003}
  w_{\alpha_1\ldots\alpha_{2J}}\; w^k\; 
\mbox{BPS}[a_1,a_2,a_3,a_4]\;.
\end{equation}
Here we have used the first factor to obtain the spin, the second 
one for the conformal dimension and the BPS factor for the 
$\mbox{SO}(8)$ quantum numbers. The unitarity bound is given by 
\begin{equation}\label{7.004}
  \ell \geq 1+J+a_1+a_2+{1\over 2}(a_3+a_4)
\end{equation}
and is saturated if $k=0$ in (\ref{7.003}). The object 
(\ref{7.003}) is short if: (i) $J\neq 0$ and $k=0$ (then it 
satisfies the intersection of (\ref{7.002}) with the BPS 
conditions); (ii) $J=0$ and $k=1$  (then it satisfies the 
intersection of (\ref{7.001}) with the BPS conditions); (iii) 
$J=0$ and $k=0$ (then it is BPS short). These results exactly 
match the classification of Ref. \cite{Minw2}.

\subsection{BPS states of $\mbox{OSp}(8/4,\mathbb{R})$}

Here we give a summary of all possible 
$\mbox{OSp}(8/4,\mathbb{R})$ BPS multiplets. Denoting the UIR's by 
\begin{equation}\label{7.39}
  {\cal D}(\ell;J;a_1,a_2,a_3,a_4)
\end{equation}
where $\ell$ is the conformal dimension, $J$ is the spin and 
$a_1,a_2,a_3,a_4$ are the $\mbox{SO}(8)$ Dynkin labels, we find 
four BPS conditions: 

\subsubsection{} 

\begin{equation}\label{8.14}
  {1\over 8}\ \mbox{BPS}:\qquad  q_\alpha^{++} =0\;.  
\end{equation}
The corresponding UIR's are: 
\begin{equation}\label{8.15}
  {\cal D}(a_1+a_2 + {1\over 2}(a_3+a_4);0;a_1,a_2,a_3,a_4)
\end{equation}
and the harmonic coset is 
\begin{equation}\label{8.16}
  {\mbox{Spin}(8)\over [\mbox{U}(1)]^4}\;.
\end{equation}
If $a_2=a_3=a_4=0$ this coset becomes 
$\mbox{Spin}(8)/\mbox{U}(4)$. 

\subsubsection{} 

\begin{equation}\label{8.17}
  {1\over 4}\ \mbox{BPS}:\qquad  q_\alpha^{++} =q_\alpha^{(++)} =0\;.  
\end{equation}
The corresponding UIR's are: 
\begin{equation}\label{8.18}
  {\cal D}(a_2 + {1\over 2}(a_3+a_4);0;0,a_2,a_3,a_4)
\end{equation}
and the harmonic coset is 
\begin{equation}\label{8.19}
  {\mbox{Spin}(8)\over [\mbox{U}(1)]^2\times \mbox{U}(2)}\;.
\end{equation}
If $a_3=a_4=0$ this coset becomes 
$\mbox{Spin}(8)/\mbox{U}(1)\times [\mbox{SU}(2)]^3$. 

\subsubsection{} 

\begin{equation}\label{8.20}
  {3\over 8}\ \mbox{BPS}:\qquad  q_\alpha^{++} =q_\alpha^{(++)} = 
q_\alpha^{[+]\{+\}} =0\;.  
\end{equation}
The corresponding UIR's are: 
\begin{equation}\label{8.21}
  {\cal D}({1\over 2}(a_3+a_4);0;0,0,a_3,a_4)
\end{equation}
and the harmonic coset is 
\begin{equation}\label{8.22}
  {\mbox{Spin}(8)\over \mbox{U}(1)\times \mbox{U}(3)}\;.
\end{equation}

\subsubsection{} \label{7.4.4}

\begin{eqnarray}
  {1\over 2}\ \mbox{BPS (type I)}:&& q_\alpha^{++} =q_\alpha^{(++)} = 
q_\alpha^{[+]\{+\}}=q_\alpha^{[+]\{\pm\}} =0\;; \label{8.23}\\ 
  {1\over 2}\ \mbox{BPS (type II)}:&& q_\alpha^{++} =q_\alpha^{(++)} = 
q_\alpha^{[+]\{+\}}=q_\alpha^{[\pm]\{+\}} =0\;. \label{8.23'} 
\end{eqnarray}
The corresponding UIR's are: 
\begin{eqnarray}
  {1\over 2}\ \mbox{BPS (type I)}:&& {\cal D}({1\over 2}a_3;0;0,0,a_3,0)\;; \label{8.24}\\ 
  {1\over 2}\ \mbox{BPS (type II)}:&& {\cal D}({1\over 2}a_4;0;0,0,0,a_4)\;. \label{8.24'} 
\end{eqnarray} 
and the harmonic coset is 
\begin{equation}\label{8.25}
  {\mbox{Spin}(8)\over \mbox{U}(4)}\;.
\end{equation}

\setcounter{equation}0 
\section{Conclusions}

Here we give a summary of the different types of BPS states which 
are realized as products of supersingletons described by 
G-analytic harmonic superfields. We shall restrict ourselves to 
the physically interesting cases of D3, $M_2$ and $M_5$ branes 
horizon geometry where only one type of such supersingletons 
appears. This construction gives rise to a restricted class of the 
most general BPS states.

\subsection{$\mbox{PSU}(2,2/4)$}

The BPS states are constructed in terms of the $N=4$ $d=4$ 
super-Yang-Mills multiplet $W^{ij}$ in three equivalent G-analytic 
realizations: 
\begin{equation}\label{8.01}
  (W^{12}(\theta_{3,4},\bar\theta^{1,2}))^{p+q+r}
(W^{13}(\theta_{2,4},\bar\theta^{1,3}))^{q+r} 
(W^{23}(\theta_{1,4},\bar\theta^{2,3}))^{r} \;.
\end{equation}

\begin{table}[h]
  \begin{center}
    \leavevmode
\label{bps4} 
    \begin{tabular}{llll}
 BPS & SU(4)  & Dimension & Harmonic space \\ \hline
  \\
 ${1\over 2}$ & (0,p,0) & p & SU(4)/S(U(2)$\times$U(2)) \\
  \\
 ${1\over 4}$ & (q,p,q) & p+2q & SU(4)/[U(1)$]^3$ \\ 
  \\
              & (q,p,q+2r) & p+2q+3r & SU(4)/[U(1)$]^3$ \\ 
 ${1\over 8}$ & (0,p,2r) & p+3r & SU(4)/U(1)$\times$U(2) \\ 
              & (0,0,2r) & 3r & SU(4)/U(3) \\ 
    \end{tabular}
  \end{center}
\end{table}

\subsection{$\mbox{OSp}(8^*/4)$}

The BPS states are constructed in terms of the $(2,0)$ $d=6$ 
tensor multiplet $W^{\{ij\}}$ in two equivalent G-analytic 
realizations: 
\begin{equation}\label{8.02}
  (W^{12}(\theta^{1,2})^{p+q}
(W^{13}(\theta^{1,3}))^{q} \;. 
\end{equation}

\begin{table}[h]
  \begin{center}
    \leavevmode
\label{bps6} 
    \begin{tabular}{llll}
 BPS & USp(4) & Dimension & Harmonic space \\ \hline
  \\
 ${1\over 2}$ & (0,p) & 2p & USp(4)/U(2) \\
  \\
 ${1\over 4}$ & (2q,p) & 2p+4q & USp(4)/[U(1)$]^2$ \\ 
              & (2q,0) & 4q & USp(4)/U(2) \\
    \end{tabular}
  \end{center}
\end{table}
\vfill\eject  

\subsection{$\mbox{OSp}(8/4,\mathbb{R})$}

The type I BPS states are constructed in terms of the $N=8$ $d=3$ 
matter multiplet $\Phi_a$ carrying an external $8_s$ $SO(8)$ 
spinor index in four equivalent G-analytic realizations: 
\begin{eqnarray}
  &&[\Phi^{+(+)[+]}
(\theta^{++,(++),[+]\{\pm\}})]^{p+q+r+s}\times \nonumber\\ 
  &&[\Phi^{+(+)[-]} (\theta^{++,(++),[-]\{\pm\}})]^{q+r+s}\times  \nonumber\\ 
  &&[\Phi^{+(-)\{+\}} (\theta^{++,(--),[\pm]\{+\}})]^{r+s}\times 
\nonumber\\ 
  &&[\Phi^{+(-)\{-\}} 
(\theta^{++,(--),[\pm]\{-\}})]^{s}\;. \label{8.03} 
\end{eqnarray}

\begin{table}[h]
  \begin{center}
    \leavevmode
\label{bps3} 
    \begin{tabular}{llll}
 BPS & SO(8)  & Dimension & Harmonic space \\ \hline
  \\
 ${1\over 2}$ & (0,0,p,0) & ${1\over 2}p$ & Spin(8)/U(4) \\
  \\
 ${1\over 4}$ & (0,q,p,0) & ${1\over 2}(p+2q)$ & Spin(8)/U(2)$\times$U(2) \\
  \\ 
 ${1\over 8}$ & (r+2s,q,p,r) & ${1\over 2}(p+2q+3r+4s)$ & 
Spin(8)/[U(1)$]^4$ \\ 
    \end{tabular}
  \end{center}
\end{table}

The type II BPS states are constructed in terms of the $N=8$ $d=3$ 
matter multiplet $\Sigma_{\dot a}$ carrying an external $8_c$ 
$SO(8)$ spinor index in four equivalent G-analytic realizations: 
\begin{eqnarray}
  &&[\Sigma^{+(+)\{+\}}
(\theta^{++,(++),[\pm]\{+\}})]^{p+q+r+s}\times \nonumber\\ 
  &&[\Sigma^{+(+)\{-\}} (\theta^{++,(++),[\pm]\{-\}})]^{q+r+s}\times  \nonumber\\ 
  &&[\Sigma^{+(-)[+]} (\theta^{++,(--),[+]\{\pm\}})]^{r+s}\times 
\nonumber\\ 
  &&[\Sigma^{+(-)[-]} 
(\theta^{++,(--),[-]\{\pm\}})]^{s}\;. \label{8.033} 
\end{eqnarray}

\begin{table}[h]
  \begin{center}
    \leavevmode
\label{bps33} 
    \begin{tabular}{llll}
 BPS & SO(8)  & Dimension & Harmonic space \\ \hline
  \\
 ${1\over 2}$ & (0,0,0,p) & ${1\over 2}p$ & Spin(8)/U(4) \\
  \\
 ${1\over 4}$ & (0,q,0,p) & ${1\over 2}(p+2q)$ & Spin(8)/U(2)$\times$U(2) \\
  \\ 
 ${1\over 8}$ & (r+2s,q,r,p) & ${1\over 2}(p+2q+3r+4s)$ & 
Spin(8)/[U(1)$]^4$ \\ 
    \end{tabular}
  \end{center}
\end{table}
\vfill\eject   

\section*{Note added} Just before submitting this paper to the hep-th archive, 
we saw a new article by P. Heslop and P.S. Howe \cite{HHowe}. It 
partially overlaps with our treatment of the $d=4$ case.

\section*{Acknowledgements}

We would like to thank V. Dobrev, L. Frappat and P. Sorba for 
enlightening discussions. E.S. is grateful to the TH Division of 
CERN for its kind hospitality. The work of S.F. has been 
supported in part by the European Commission TMR programme 
ERBFMRX-CT96-0045 (Laboratori Nazionali di Frascati, INFN) and by 
DOE grant DE-FG03-91ER40662, Task C.

\vfill\eject


\end{document}